\documentclass[1p,preprint]{elsarticle}
\usepackage{graphicx,amsmath,amssymb,amsfonts,bm}
\usepackage[figuresright]{rotating}

\begin{document}
\begin{frontmatter}

\title{Measurement of the Nucleon Structure Function $F_2$
in the Nuclear Medium and Evaluation of its Moments}


\newcommand*{\ANL}{Argonne National Laboratory, Argonne, Illinois 60441}
\newcommand*{\ANLindex}{1}
\newcommand*{\ASU}{Arizona State University, Tempe, Arizona 85287-1504}
\newcommand*{\ASUindex}{2}
\newcommand*{\CSUDH}{California State University, Dominguez Hills, Carson, CA 90747}
\newcommand*{\CSUDHindex}{3}
\newcommand*{\CANISIUS}{Canisius College, Buffalo, NY}
\newcommand*{\CANISIUSindex}{4}
\newcommand*{\CMU}{Carnegie Mellon University, Pittsburgh, Pennsylvania 15213}
\newcommand*{\CMUindex}{5}
\newcommand*{\CUA}{Catholic University of America, Washington, D.C. 20064}
\newcommand*{\CUAindex}{6}
\newcommand*{\SACLAY}{CEA, Centre de Saclay, Irfu/Service de Physique Nucl\'eaire, 91191 Gif-sur-Yvette, France}
\newcommand*{\SACLAYindex}{7}
\newcommand*{\CNU}{Christopher Newport University, Newport News, Virginia 23606}
\newcommand*{\CNUindex}{8}
\newcommand*{\UCONN}{University of Connecticut, Storrs, Connecticut 06269}
\newcommand*{\UCONNindex}{9}
\newcommand*{\EDINBURGH}{Edinburgh University, Edinburgh EH9 3JZ, United Kingdom}
\newcommand*{\EDINBURGHindex}{10}
\newcommand*{\FU}{Fairfield University, Fairfield CT 06824}
\newcommand*{\FUindex}{11}
\newcommand*{\FIU}{Florida International University, Miami, Florida 33199}
\newcommand*{\FIUindex}{12}
\newcommand*{\FSU}{Florida State University, Tallahassee, Florida 32306}
\newcommand*{\FSUindex}{13}
\newcommand*{\GWU}{The George Washington University, Washington, DC 20052}
\newcommand*{\GWUindex}{14}
\newcommand*{\ISU}{Idaho State University, Pocatello, Idaho 83209}
\newcommand*{\ISUindex}{15}
\newcommand*{\INFNFR}{INFN, Laboratori Nazionali di Frascati, 00044 Frascati, Italy}
\newcommand*{\INFNFRindex}{16}
\newcommand*{\INFNGE}{INFN, Sezione di Genova, 16146 Genova, Italy}
\newcommand*{\INFNGEindex}{17}
\newcommand*{\INFNRO}{INFN, Sezione di Roma Tor Vergata, 00133 Rome, Italy}
\newcommand*{\INFNROindex}{18}
\newcommand*{\ORSAY}{Institut de Physique Nucl\'eaire ORSAY, Orsay, France}
\newcommand*{\ORSAYindex}{19}
\newcommand*{\ITEP}{Institute of Theoretical and Experimental Physics, Moscow, 117259, Russia}
\newcommand*{\ITEPindex}{20}
\newcommand*{\JMU}{James Madison University, Harrisonburg, Virginia 22807}
\newcommand*{\JMUindex}{21}
\newcommand*{\KNU}{Kyungpook National University, Daegu 702-701, Republic of Korea}
\newcommand*{\KNUindex}{22}
\newcommand*{\LPSC}{LPSC, Universite Joseph Fourier, CNRS/IN2P3, INPG, Grenoble, France}
\newcommand*{\LPSCindex}{23}
\newcommand*{\UNH}{University of New Hampshire, Durham, New Hampshire 03824-3568}
\newcommand*{\UNHindex}{24}
\newcommand*{\NSU}{Norfolk State University, Norfolk, Virginia 23504}
\newcommand*{\NSUindex}{25}
\newcommand*{\OHIOU}{Ohio University, Athens, Ohio  45701}
\newcommand*{\OHIOUindex}{26}
\newcommand*{\ODU}{Old Dominion University, Norfolk, Virginia 23529}
\newcommand*{\ODUindex}{27}
\newcommand*{\RPI}{Rensselaer Polytechnic Institute, Troy, New York 12180-3590}
\newcommand*{\RPIindex}{28}
\newcommand*{\URICH}{University of Richmond, Richmond, Virginia 23173}
\newcommand*{\URICHindex}{29}
\newcommand*{\ROMAII}{Universita' di Roma Tor Vergata, 00133 Rome, Italy}
\newcommand*{\ROMAIIindex}{30}
\newcommand*{\ROMAIII}{INFN, Sezione di ROMA III, 00146 Rome, Italy}
\newcommand*{\ROMAIIIindex}{31}
\newcommand*{\MSU}{Skobeltsyn Nuclear Physics Institute, Skobeltsyn Nuclear Physics Institute, 119899 Moscow, Russia}
\newcommand*{\MSUindex}{32}
\newcommand*{\SCAROLINA}{University of South Carolina, Columbia, South Carolina 29208}
\newcommand*{\SCAROLINAindex}{33}
\newcommand*{\JLAB}{Thomas Jefferson National Accelerator Facility, Newport News, Virginia 23606}
\newcommand*{\JLABindex}{34}
\newcommand*{\UNIONC}{Union College, Schenectady, NY 12308}
\newcommand*{\UNIONCindex}{35}
\newcommand*{\UTFSM}{Universidad T\'{e}cnica Federico Santa Mar\'{i}a, Casilla 110-V Valpara\'{i}so, Chile}
\newcommand*{\UTFSMindex}{36}
\newcommand*{\GLASGOW}{University of Glasgow, Glasgow G12 8QQ, United Kingdom}
\newcommand*{\GLASGOWindex}{37}
\newcommand*{\VIRGINIA}{University of Virginia, Charlottesville, Virginia 22901}
\newcommand*{\VIRGINIAindex}{38}
\newcommand*{\WM}{College of William and Mary, Williamsburg, Virginia 23187-8795}
\newcommand*{\WMindex}{39}
\newcommand*{\YEREVAN}{Yerevan Physics Institute, 375036 Yerevan, Armenia}
\newcommand*{\YEREVANindex}{40}
 
\newcommand*{\NOWCUA}{Catholic University of America, Washington, D.C. 20064}
\newcommand*{\NOWLANL}{Los Alamos National Laborotory, New Mexico, NM}
\newcommand*{\NOWJLAB}{Thomas Jefferson National Accelerator Facility, Newport News, Virginia 23606}
\newcommand*{\NOWGWU}{The George Washington University, Washington, DC 20052}
\newcommand*{\NOWCNU}{Christopher Newport University, Newport News, Virginia 23606}
\newcommand*{\NOWORSAY}{Institut de Physique Nucl\'eaire ORSAY, Orsay, France}
\newcommand*{\NOWEDINBURGH}{Edinburgh University, Edinburgh EH9 3JZ, United Kingdom}
\newcommand*{\NOWWM}{College of William and Mary, Williamsburg, Virginia 23187-8795}

\author[toINFNGE,toMSU]{M.~Osipenko\corref{cor1}}
\author[toINFNGE]{G.~Ricco}
\author[toROMAIII]{S.~Simula}
\author[toINFNGE]{M.~Ripani}
\author[toINFNGE]{M.~Taiuti}
\author[toODU]{K.P.~Adhikari}
\author[toODU]{M.J.~Amaryan}
\author[toINFNGE]{M.~Anghinolfi}
\author[toJLAB,toINFNFR]{H.~Avakian}
\author[toVIRGINIA]{H.~Baghdasaryan}

% ...
\address[toINFNFR]{\INFNFR} 
\address[toINFNGE]{\INFNGE} 
% ...
\address[toODU]{\ODU} 
% ...
\address[toROMAIII]{\ROMAIII} 
\address[toMSU]{\MSU} 
% ...
\address[toJLAB]{\JLAB} 
% ...
\address[toVIRGINIA]{\VIRGINIA} 
% ...

\fntext[toNOWCUA]{Current address: Washington, D.C. 20064 }
\fntext[toNOWLANL]{Current address: New Mexico, NM }
\fntext[toNOWJLAB]{Current address: Newport News, Virginia 23606 }
\fntext[toNOWGWU]{Current address: Washington, DC 20052 }
\fntext[toNOWCNU]{Current address: Newport News, Virginia 23606 }
\fntext[toNOWORSAY]{Current address: Orsay, France }
\fntext[toNOWEDINBURGH]{Current address: Edinburgh EH9 3JZ, United Kingdom }
\fntext[toNOWWM]{Current address: Williamsburg, Virginia 23187-8795 }
\cortext[cor1]{Corresponding author: osipenko@ge.infn.it}


\begin{abstract}
We report on the measurement of inclusive electron scattering off a carbon target 
performed with CLAS at Jefferson Laboratory.
A combination of three different beam energies 1.161, 2.261 and 4.461 GeV
allowed us to reach an invariant mass of the final-state hadronic system $W\approx 2.4$~GeV
with four-momentum transfers $Q^2$ ranging from $0.2$ to $5$~(GeV/c)$^2$.
These data, together with previous measurements of the inclusive electron
scattering off proton and deuteron, which cover a similar continuous two-dimensional
region of $Q^2$ and Bjorken variable $x$, permit the study of nuclear modifications
of the nucleon structure.
By using these, as well as other world data, we evaluated the $F_2$ structure function
and its moments.
Using an OPE-based twist expansion, we studied the $Q^2$-evolution of the moments,
obtaining a separation of the leading-twist and the total higher-twist terms.
The carbon-to-deuteron ratio of the leading-twist contributions to the $F_2$ moments
exhibits the well known EMC effect, compatible with that discovered previously
in $x$-space. The total higher-twist term in the carbon nucleus appears,
although with large systematic uncertainties,
to be smaller with respect to the deuteron case for $n < 7$, suggesting
partial parton deconfinement in nuclear matter. We speculate that the spatial
extension of the nucleon is changed when it is immersed in the nuclear medium.
\end{abstract}

\end{frontmatter}

\section{\label{sec:introduction}Introduction}
Inclusive electron scattering off a nucleus,
$e(k) + A(P) \to e^\prime (k^\prime) + X$,
may be expressed, under the one photon exchange approximation, in terms of 
the total
absorption of a virtual photon by the nucleus: $\gamma^*(q=k-k^\prime) + A(P) \to X$.
At sufficiently large squared four-momentum transfer, $Q^2=-q^2$,
the coherent electron-nucleus 
scattering contribution is negligible, facilitating a description 
of the scattering in terms of individual nucleons in the nuclear matter.
The inclusive electron-nucleon scattering cross section is
related to the parton momentum distribution along the virtual photon (longitudinal)
direction.
This structure, however, depends on the environment in which the nucleon is embedded.
In particular, the nuclear medium environment interferes with the nucleon internal structure.
The experimental evidence for such nuclear modifications in
the Deep Inelastic Scattering (DIS) regime was found by
the European Muon Collaboration (EMC)~\cite{EMC-effect}.
Other similar observations came from intermediate energy experiments
that observed changes in the nucleon and nucleon excited state form-factors~\cite{res_suppress}.
In order to study the transition of the nuclear modifications
from the DIS regime (EMC effect) to the larger spacial scale region
(the nucleon and its excited state form-factors),
the structure function $F_2$ is calculated from the cross section and used
to perform a Mellin transform to moments $M^{CN}$.
The Mellin transform allows to express the observable from Minkowski $x$-space
($x$-representation of $F_2(x)$)
in Euclidean space of moments ($n$-representation of $M^{CN}_n$),
where Lattice QCD method works.

The Cornwall-Norton moments $M^{CN}_n$ 
of the nucleon (in the nucleus with mass $M_A$) structure functions are defined as:
\begin{equation}
M^{CN}_n(Q^2)=\int_0^{M_A/M} dx x^{(n-2)}F_2(x,Q^2),~~~\mbox{$n\ge2$, $n$ even},
\label{eq:CWMoment}
\end{equation}
\noindent where $M$ is the nucleon mass, $x=-A q^2/2Pq$ is the Bjorken variable per nucleon
and $n$ is the order of the moment.
Quantum Chromodynamics (QCD) predicts the $Q^2$-evolution of these moments
for each given $n$~\cite{Roberts}.
Moreover, the DIS component of these moments
and the contribution of the nucleon and the nucleon excited state form-factors
have different $Q^2$ behaviors~\cite{resOPE}. 
Therefore, the study of
the $Q^2$ dependence of the structure function moments is important
for understanding the nucleon modifications in the nuclear medium.

The identification of these nuclear modifications to the nucleon structure
in inclusive electron scattering is complicated by the presence of
other nuclear effects~\cite{Arneodo_review}. The most important effect
comes from the motion of the nucleon in the nucleus, known as Fermi motion.
There are also other effects that cannot be reduced to the total absorption
of the virtual photon by the nucleon. These are shadowing,
Final State Interactions (FSIs) and Meson Exchange Currents (MECs).
However, the latter ones are typically smaller
and can be localized by their kinematic regions of influence.

The study of nuclear modifications can be performed by comparing 
the nucleon structure of a free nucleon to one bound in the nuclear medium.
To this end, the proton and deuteron structure function moments
were previously measured in Refs.~\cite{osipenko_f2p,osipenko_f2d}.
The present article maintains the same framework of these two previous measurements
in order to avoid additional bias due to the analysis technique.

\begin{figure}
\begin{center}
\includegraphics[bb=1cm 6cm 20cm 23cm, scale=0.4]{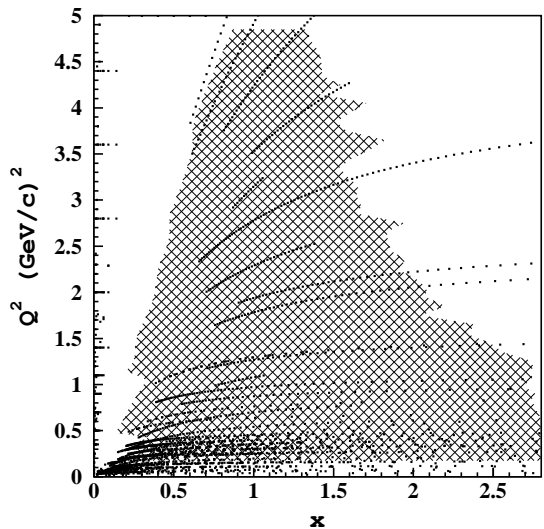}
\caption{\label{fig:xandQ2Domain} Experimental data on the carbon structure
function $F_2(x,Q^2)$ used for the moment evaluation in the CLAS
kinematic region.
The points show the world data from
Refs.~\cite{gomez94,baran88,day93,desy73,hall_c,kiev69,oconnell84,saclay83,sealock89,yerevan88,Harvard71,Arrington95,Whitney74,bcdms94,bcdms,cern81,emc,cornell76,daresbury79,desy77,fermilab81,e665,emc88_p1,emc88_p2,emc89,nmc95}.
The shaded area shows the CLAS data region.}
\end{center}
\end{figure}

Similar studies were performed in the past in Ref.~\cite{Ricco2}
and recently repeated by the Hall~C Collaboration at Jefferson Lab~\cite{hallc_nucl_moms},
and reported contradictory results.
In this paper, we increase the precision of the previous studies by
a new measurement of unpolarized inclusive electron
scattering on carbon, performed with the CLAS detector in Hall~B at Jefferson Lab.
Like in the case of the proton and deuteron measurements,
the new data cover a wide continuous two-dimensional region
in $x$ and $Q^2$ as shown in Fig.~\ref{fig:xandQ2Domain}.

In Section~\ref{sec:Moments} we describe the QCD picture of the $F_2$ moments.
In Section~\ref{sec:DataAnalysis} we discuss some data analysis details.
For further information we refer to Refs.~\cite{osipenko_f2p,osipenko_f2d},
from which many procedures are adopted.
Finally, in Sections~\ref{sec:Discussion} and \ref{sec:Comparison}
we present our main results.

\section{QCD Analysis of Nucleon Structure}\label{sec:Moments}
QCD-based descriptions of inclusive lepton scattering can 
require Operator Product Expansion (OPE) machinery~\cite{Roberts}.
The OPE expresses the structure function moments in a power series:
\begin{equation}
M^{CN}_n(Q^2)=\sum_{\tau=2k}^{\infty}E_{n\tau}(\mu_r,\mu_f,\alpha_S)
O_{n\tau}(\mu_f)\biggl(\frac{\mu^2}{Q^2}\biggr)^{\frac{1}{2}(\tau-2)},
\label{eq:i_m1}
\end{equation}
\noindent
where $k=1,2,...,\infty$, $\mu_f$ ($\mu_r$) is the factorization (renormalization)
scale$^1$\footnotetext[1]{We are working in the Soft Gluon Re-summation (SGR)
scheme~\cite{SGR}, where $\mu_f^2=\mu_r^2=Q^2$.} and $\mu$ is an arbitrary reference scale.
$O_{n\tau}(\mu_f)$ is the reduced matrix element of the local operators
with definite spin $n$ and twist $\tau$ (dimension minus spin).
This quantity is related to the partonic structure of the target.
$E_{n\tau}(\mu_r,\mu_f,\alpha_S)$ is a dimensionless coefficient function.
For sufficiently large $Q^2$ and therefore small running coupling constant $\alpha_s(Q^2)$,
it can be analytically calculated in perturbative QCD (pQCD).

The first term in this expansion (the leading twist):
\begin{equation}
\eta_n(Q^2)=E_{n 2}(\mu_r,\mu_f,\alpha_S)O_{n 2}(\mu_f),
\label{eq:lt_def}
\end{equation}
\noindent can be completely
evaluated in QCD. The operator matrix elements $O_{n 2}$, given by
the expectation values of the parton momentum distributions in the nucleon,
are calculated in Lattice QCD~\cite{LatticeF2} for $n=2,4$.
The coefficient functions are calculated in pQCD~\cite{pqcdNNLO}
at $\alpha_S^2$ precision, including resummation of specific effects in
the large-$x$ region to $\ln^2{(\alpha_S)}$ precision~\cite{SGR,SIM00}.
This term is responsible for the DIS inclusive cross section
and its modification in nuclei gives rise to the EMC effect.

Further terms are far less known, but can be studied phenomenologically
by exploiting the general form of their $Q^2$-dependence.
The separation of individual twists related to specific QCD operators
is not possible without knowledge of the corresponding coefficient
functions, but the total higher-twist term:
\begin{equation}
HT_n(Q^2)=\frac{\mu^2}{Q^2}
\sum_{\tau=4}^{\infty}E_{n\tau}(\mu_r,\mu_f,\alpha_S)
O_{n\tau}(\mu_f)\biggl(\frac{\mu^2}{Q^2}\biggr)^{\frac{1}{2}(\tau-4)}
\label{eq:ht_def}
\end{equation}
\noindent contribution can still be
analyzed. This total higher-twist term is responsible for the
multi-parton correlations in the nucleon and, therefore, it is
connected also to the nucleon elastic form-factors and to
the transition form-factors to the nucleon excited states.
If there are nucleon modifications of these properties
that are important at lower $Q^2$, higher twists can be used to investigate them. 

In summary, a comparison of the leading-twist moments of the
nucleon structure function in complex nuclei to that in the deuteron
allows a study of the EMC effect in the moment space ($M_n(Q^2)$). This allows a direct
comparison to Lattice QCD calculations of the nucleon modified by the presence of the medium.
A similar comparison between the total higher-twist moments in deuteron and nuclei
allows study of the nuclear modifications of confinement and multi-parton correlations.

\section{Data Analysis}\label{sec:DataAnalysis}
The data were collected at Jefferson Lab in Hall B with the CEBAF
Large Acceptance Spectrometer (CLAS)
using a solid, $0.18$~g/cm$^2$ thick, carbon target 
during the electron beam running period in April-May 1999.
The average luminosity was
6$\times$10$^{33}$ cm$^{-2}$s$^{-1}$.
Data were recorded using electron beam energies 
$E_0=$~1.161, 2.261 and 4.461 GeV to maximize the $Q^2$ and $x$ kinematics range.
The accumulated statistics at the three energies allows a measurement of the inclusive cross section with a 
statistical uncertainty ($\le$ 5\%) in $x$ and $Q^2$ intervals of 
$\Delta x=$0.009 and $\Delta Q^2=$0.05-0.1 (GeV/c)$^2$.

CLAS is a magnetic spectrometer~\cite{CLAS_paper} based on a six-coil
torus magnet whose field is primarily oriented along the azimuthal
direction. The sectors, located between the magnet coils, are individually
instrumented to form six independent magnetic spectrometers. The particle
detection system includes Drift Chambers (DC) for track
reconstruction~\cite{dc},
Scintillation Counters (SC) for time-of-flight measurements~\cite{sc},
Cherenkov Counters (CC) for electron identification~\cite{cc}, and
Electromagnetic Calorimeters (EC) to measure neutrals and to improve
electron-pion separation~\cite{ec}. The EC detectors, which have
a granularity defined by triangular cells in a plane perpendicular
to the incoming particles, are used to study the shape of the
electromagnetic shower and
are longitudinally divided into two parts with the inner part acting as
a pre-shower.

Charged particles can be detected and identified for momenta
down to 0.2 (GeV/c) and for polar angles between 8$^\circ$ and 142$^\circ$.
The CLAS superconducting coils limit the acceptance 
for charged hadrons from about 80\%
at $\theta=90^{\circ}$ to about 50\% at forward angles ($\theta=20^{\circ}$).
The total angular acceptance for electrons is about 1.5 sr.
Electron momentum resolution is a function of the scattered electron angle
and varies from 0.5\% for $\theta \leq 30^{\circ}$ up to
1-2\% for $\theta > 30^{\circ}$. The angular resolution
is approximately constant, approaching 1~mrad for polar and 4~mrad
for azimuthal angles: the resolution for the momentum transfer
ranges therefore from 0.2 up to 0.5 \%. The scattered electron missing mass ($W$)
resolution was estimated to be 2.5 MeV for a beam energy less than 3 GeV and
about 7 MeV for larger energies.
To study all possible multi-particle states, the acquisition trigger
was configured to require at least one electron candidate in any of
the sectors, where an electron candidate was defined as
the coincidence of a signal in the EC and Cherenkov modules
for any one of the sectors.

The data analysis procedure has been described in detail in
Refs.~\cite{osipenko_f2p,osipenko_f2d}.
Therefore, in this article we focus on changes and improvements in the analysis.
The most important improvements, leading to a significant reduction
of the estimated systematic uncertainties relative to those of
Refs.~\cite{osipenko_f2p,osipenko_f2d}, are described in the following sections.

\subsection{Generic Procedures}\label{sec:gen_proc}
\begin{figure}
\begin{center}
\includegraphics[bb=5cm 5cm 16cm 23cm, scale=0.4]{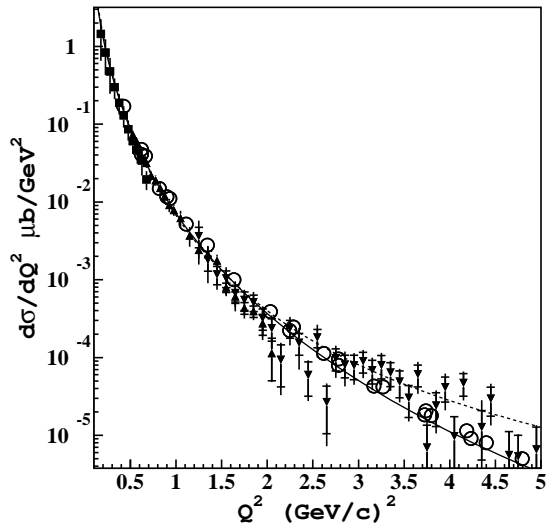}
\caption{\label{fig:h_el} Hydrogen elastic scattering cross section
extracted by means of CH$_2$-C subtraction for three different beam energies:
full squares - 1.161 GeV,
full triangles - 2.261 GeV,
reversed full triangles - 4.461 GeV.
The inner error bars show statistical uncertainties, while the outer error bars
represent the statistical and systematic uncertainties summed in quadrature.
The cross sections are compared with the data from Ref.~\cite{Christy04} (empty circles)
and with parametrization based on dipole form-factors (dashed line)
including also radiative corrections (solid line).}
\end{center}
\end{figure}

Phenomenological corrections to the reconstructed charged particle momentum were 
applied to compensate for small drift chamber misalignments and torus magnetic field map 
inaccuracies.
These corrections were determined by subtracting the missing mass spectrum ($W$) 
measured using a solid carbon target from a measurement using CH$_2$ in order to extract 
the proton elastic scattering response.
The scattered electron momentum measurement was shifted for
the entire $W$ spectrum in such a way that the
elastic scattering peak was at the correct position of $W=\mbox{938 MeV/c}^2$. 
This was accomplished by making a fit
of the hydrogen elastic peak distributions for a few azimuthal angle bins
inside each sector under the assumption that the polar angle was
measured precisely.
These corrections were applied for
each sector separately and improved the elastic missing mass resolution by 15\%.
For the highest beam energy data set, the statistics from hydrogen
were insufficient to determine momentum corrections.
However, the Fermi motion at these $Q^2$ values leads to a very smooth cross
section having no peaks or other structures and, therefore, makes it relatively insensitive
to possible small systematic uncertainties in the measured momentum.
To check the absolute normalization of our data we subtracted the carbon data
from CH$_2$ yield and obtained the cross section of the scattering on hydrogen.
The elastic scattering cross section was selected and compared
to the previous data from Ref.~\cite{Christy04} and to the dipole parametrization.
The comparison, shown in Fig.~\ref{fig:h_el}, exhibits fairly good agreement,
although with large systematic uncertainties, related to the small fraction
of hydrogen events in the total CH$_2$ yield.
%

Although the results of the present analysis use
realistic Monte Carlo simulations of the CLAS detector, such simulations have
limited reliability at the edges of the detector's sensitive volume.
There are also small regions of the CLAS detector that were very inefficient
and, in some cases, were malfunctioning.
These regions were eliminated using kinematic cuts to simplify the analysis.
The kinematic cuts were defined as functions of the measured 
scattered electron momentum and angle
to establish fiducial volumes with a uniform detector efficiency in the polar angle $\phi$.
These cuts along the
edges of the acceptance removed about 50\% of the recorded data.

\subsection{Electron Identification}\label{sec:ElPid}
Electrons were identified in CLAS by a coincidence between
signals from the Electromagnetic Calorimeter (EC) and the Cherenkov Counter (CC).
Furthermore, only events with a reconstructed Drift Chamber (DC)
negatively charged particle track that matches these hits were selected.

The CLAS Cherenkov Counter (CC) inefficiency does not exceed 2\% within the 
fiducial regions of the CLAS acceptance~\cite{osipenko_f2p,osipenko_f2d}.
However, the electron-pion separation does not
work at the same level of precision and some pion contamination shows up in
the electron candidate sample. This is especially important at high
beam energies (e.g. for the 4.461 GeV data set), where the number of low momentum pions
is significant.

\begin{figure}[!h]
\begin{center}
\includegraphics[bb=1cm 6cm 20cm 23cm, scale=0.4]{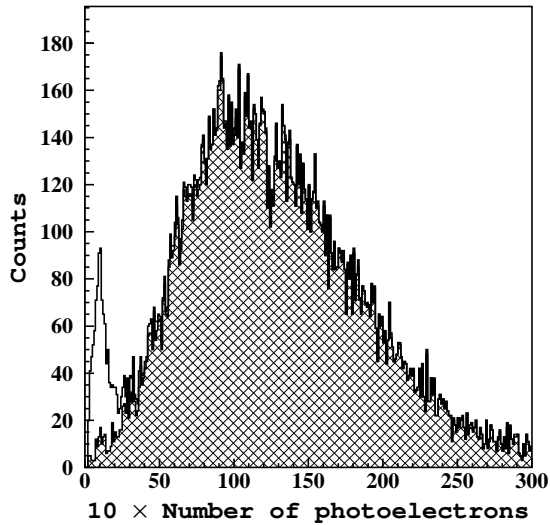}
\caption{\label{fig:after_cut} The spectrum of photoelectrons measured
in one sector of the CC at $\nu=3.4$ GeV (beam energy 4.461 GeV).
The hatched area represents the CC spectrum after applying the matching
procedure described in the text.}
\end{center}
\end{figure}

The pion contamination in the electron candidate sample appears as a 
single photoelectron peak in the 
measured CC spectra (see Fig.~\ref{fig:after_cut}). 
The main source of this contamination is the result of an accidental coincidence between a 
pion produced by quasi-real photoproduction with random noise in one of the 36 CC PMTs.
This coincidence takes place due 
to the loose matching between the CC hit and the track within 
one sector of CLAS.
A procedure was developed in Ref.~\cite{CCmatch} to geometrically and temporally
match a CC hit with a track seen in the DC and a hit in the SC.
The resulting Cherenkov distribution is shown in Fig.~\ref{fig:after_cut}.
The reconstruction algorithm eliminated the pion contamination in
the single-photoelectron peak, which in the worst case could reach 20\%.

\begin{figure}
\begin{center}
\includegraphics[bb=5cm 5cm 16cm 23cm, scale=0.4]{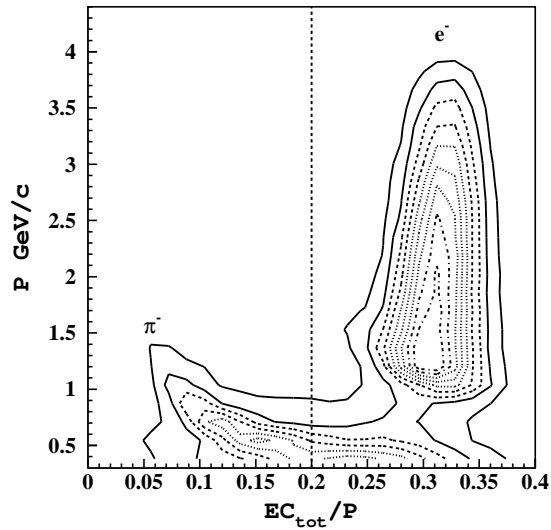}
\caption{\label{fig:el_pid}Electron-pion separation in the Electromagnetic Calorimeter (EC) of CLAS:
the left peak at small released energy fraction is due to pions and
the right peak at $EC_{tot}/\mathbf{P} \sim 0.3$ represents the electron signal.
The dashed line shows the cut applied to the data.}
\end{center}
\end{figure}

The Electromagnetic Calorimeter (EC) is used to separate pions from electron candidates
with momenta above the pion threshold (2.7 GeV/c).
Electrons entering in the EC
release $\sim 30$\% of their energy in the sensitive volume on average,
while pion losses are constant (see Fig.~\ref{fig:el_pid}).
This EC property was exploited at large particle momenta for
electron-pion separation by selecting particles with an energy
fraction released in the EC above 20\%. More details on this procedure
can be found in Ref.~\cite{osipenko_f2p}.
Furthermore, pions just above the Cherenkov
threshold produce less Cherenkov light with respect to electrons of
the same momenta and, therefore, can be removed by a cut on the
number of photoelectrons measured in the CC.

\subsection{Background}
The creation of $e^+e^-$ pairs
by real photons or through hadronic decays and the scattering of electrons 
from materials other than the target are the dominant sources of background electrons.

\subsubsection{$e^+e^-$ pair production}
The main source of $e^+e^-$ pairs entering the CLAS 
acceptance is from $\pi^0$ production, followed by either Dalitz decay to $\gamma e^+e^-$
or $\pi^0$ decay to $\gamma \gamma$, where  one of the photons converts to an $e^+e^-$ pair.  
A model using a Wiser fit to inclusive pion photoproduction was used to effectively 
reproduce the positron cross sections observed  in a previous CLAS experiment~\cite{BostedPP}.
The above model was used to estimate the $e^+e^-$ contribution 
to the measured inclusive cross section. The value of the contribution was assumed
to be equal to the ratio of the inclusive $e^+$ production cross
section over the inclusive nuclear scattering 
parametrization~\cite{Ricco2,Simula_nucl} including radiative processes
(tail from elastic peak, bremsstrahlung and Schwinger correction):
\begin{equation}\label{eq:d_pp1}
F_{e^+e^-}(E_0,x,Q^2)=\frac{
\sigma^M_{rad}(E_0,x,Q^2)}{
\sigma^M_{rad}(E_0,x,Q^2)+
\sigma_{e^+}(E_0,x,Q^2)}  ~~,
\end{equation}
\noindent where $\sigma_{e^+}$ is the inclusive $e^+$ production cross
section, equal to the integral of the $\pi^0$ quasi-real electroproduction cross section
folded with its decay probability and integrated over the allowed kinematic range,
and $\sigma^M_{rad}$ is the inclusive nuclear scattering parametrization~\cite{Ricco2,Simula_nucl}
folded with radiative processes, as described in Ref.~\cite{osipenko_f2p}.

In this parametrization the $\pi^0$ quasi-real electroproduction cross section is given by:
\begin{equation}\label{eq:d_pp4}
\frac{d\sigma_{ep\to hX}}{dp_h d\Omega_h}=\frac{P_h^2}{E_h}\int_{E_\gamma^{min}}^{E_0}\frac{dE_\gamma}{E_\gamma}
\Bigl\{\frac{t_{ext}}{2}+t_{int}\Bigr\}
E_h\frac{d^3 \sigma_{\gamma p\to hX}}{d^3 p_h} \Bigl[e^{C_1 M_L} e^{C_2 p_T^2/E_h}\Bigr] ~~,
\end{equation}
\noindent where $\sigma_{\gamma p\to hX}$ is the parametrized photoproduction cross section,
$P_h$ and $E_h$ are the pion momentum and energy, $p_T$ is the pion transverse momentum
with respect to the photon direction, $E_0$ is the electron beam energy, $E_\gamma$ is the exchanged photon energy,
$t_{int}$ and $t_{ext}$ are the internal and external radiator thicknesses
and $M_L=\sqrt{p_T^2-m_h^2}$, with $m_h$ being the pion mass. $C_1$ and $C_2$ are the Wiser fit parameters.

The external radiator thickness is small compared to the second term, $t_{ext}=0.0053$ radiation lengths.
The internal radiator length used to estimate the $e^+$ rate (previously assumed to be 5\%)
was evaluated according to the Weizsacker-Williams formula~\cite{Brodsky_epem}:
\begin{equation}\label{eq:d_pp3}
t_{int}=\frac{\alpha}{2}\Biggl[
\frac{E_0^2+E^{\prime 2}}{E_0^2}\Bigl(\log{\frac{E_0}{m_e}}-\frac{1}{2}\Bigr)+
\frac{E_\gamma^2}{2E_0^2}\Bigl(\log{\frac{2E^\prime}{E_\gamma}}+1\Bigr)+
\frac{(E_0+E^\prime)^2}{2E_0^2}\log{\frac{2E^\prime}{E_0+E^\prime}}
\Biggr] ~~,
\end{equation}
\noindent where $E^\prime$ is the scattered electron energy and $m_e$ is the electron mass.

\begin{figure}[!h]
\begin{center}
\includegraphics[bb=1cm 6cm 20cm 23cm, scale=0.4]{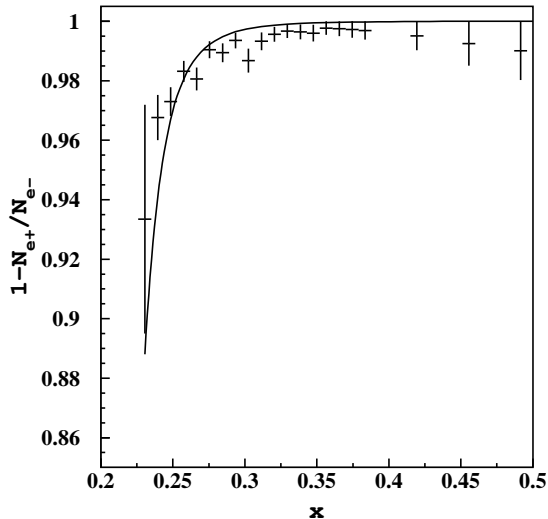}
\caption{\label{fig:epem} The contribution of $e^+e^-$ pair
production events in the inclusive cross section at $Q^2=1.55$ (GeV/c)$^2$.
The numbers of positrons $N_{e^+}$ and electrons $N_{e^-}$
were obtained from different runs with opposite CLAS torus magnetic fields,
allowing for a cancellation of the acceptance and efficiency in the ratio.
The points show the measured quantity $1-N_{e^+}/N_{e^-}$,
which represents the ratio of the number of electrons inelastically scattered off carbon
to the total number of measured electrons.
The curve represents the calculations from Eq.~\ref{eq:d_pp1}.}
\end{center}
\end{figure}

The largest $e^+e^-$ contribution is less than 9\% of the inclusive cross section
at $x$ of 0.2 as shown in Fig.~\ref{fig:epem} and only impacts the data 
taken at 2.261 and 4.461 GeV beam energies, where such low values of $x$ are accessible.

\subsubsection{Target Cell Contribution}
\begin{figure}[!h]
\begin{center}
\includegraphics[bb=1cm 6cm 20cm 23cm, scale=0.4]{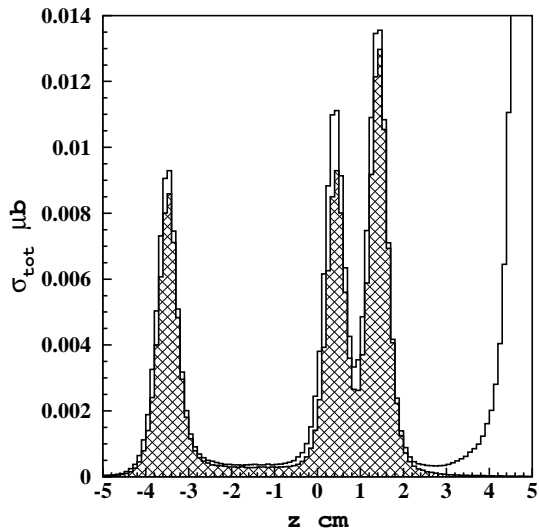}
\caption{\label{fig:empty_target} $Z$-vertex distribution for
the carbon target in (open histogram), and carbon target out (hatched histogram)
at $E_0=1.161$ GeV.
The large peak that can be partially seen on the right side of the plot is due to
scattering off the carbon plate, while the three smaller peaks are due to the empty target windows.}
\end{center}
\end{figure}

The target wall contribution to the measured cross section has been removed using 
empty target data in bins of the energy $E_0$, the Bjorken scaling variable $x$
and $Q^2$ such that:
\begin{equation}\label{eq:d_et1}
N_C(E_0,x,Q^2)=N_{full}(E_0,x,Q^2)-
\frac{Q_{tot}^{full}}{Q_{tot}^{empty}}
N_{empty}(E_0,x,Q^2) ~~,
\end{equation}
\noindent where $N_{full}$, $N_{empty}$ and $N_C$ are the number of events
in the full target data, the empty target data and the number of events due to scattering
off the carbon target, respectively.
$Q_{tot}^{full}$ and $Q_{tot}^{empty}$ 
represent the charge accumulated in full and empty target runs as measured 
by a Faraday Cup (FC) gated by the data acquisition live time.
Any inaccuracy of the $Z$-vertex reconstruction (see Fig.~\ref{fig:empty_target})
did not affect the extracted
cross section because this information was not used
in the evaluation procedure.

Multiple scattering and elastic electron-nucleus scattering at 
the lowest beam energy of $E_0=1.161$ GeV caused the incident electron beam 
to diverge after 
passing through a full target such that a small fraction of the beam was no longer 
within the FC acceptance. This resulted in charge loss.
A precise experimental determination of the charge loss
was performed using the measured cross section of electron scattering
by the empty target walls (Mylar/Aluminum windows of the empty cell for liquids).
Electrons scattered by the empty cell walls were selected according to their
reconstructed $Z$-vertex shown in Fig.~\ref{fig:empty_target}.
The evident enhancement of the full target
data is due to the FC charge loss. The overall normalization factor was estimated
by selecting the region of $Z$ from -6 cm to 3 cm in both empty and full target
data and evaluating the ratio:
\begin{equation}\label{eq:d_cl1}
F_{norm}=\frac{\int_{z=-6 cm}^{z=3 cm}
\sigma_{tot}^{full}dz}{\int_{z=-6 cm}^{z=3 cm} \sigma_{tot}^{empty}dz} ~~,
\end{equation}
\noindent where $\sigma_{tot}^{full,empty}(Z)=\frac{1}{\rho \frac{N_A}{M_A} L Q_{tot}}N_{events}^{full,empty}(Z)$
is the total cross section for the target corresponding to a given $Z$ interval
(for definition of variables see Section~\ref{sec:d_sf}).
The correction factor was found to be $F_{norm}^{1.161 GeV}=1.1459$ with
negligible statistical uncertainty. Selecting only the first empty target cell
window using the cut on $Z$ from -6 cm to -1 cm gives the same ratio within 0.2\%.
The corresponding $Z$-vertex distribution shows good agreement
between the empty and full target cross sections using the above correction factor $F_{norm}$.
Numerical estimates for $E_0=4.461$ GeV data
give the ratio $F_{norm}^{4.461 GeV}=1.0006 \pm 0.006$ for the $Z$-cut from -10 cm to -2.5 cm.
A systematic uncertainty of  3\% was assigned to the FC charge loss
correction for the $E_0=1.161$ GeV data set based on the above analysis.

\subsection{Acceptance and Efficiency Evaluation}
Determination of the acceptance and efficiency 
corrections was based entirely on the Monte Carlo (MC) simulations developed for CLAS.
Moreover, the systematic uncertainties of these corrections
were estimated from a comparison of MC simulations with
experimental data using a realistic model in the event generator.
In short, the procedure was the following:
we generated events with the event generator describing
elastic, quasi-elastic and inelastic $eC$-scattering processes including
radiative corrections. These events then were processed with
the GEANT-based CLAS software simulating
the detector response. After that the standard CLAS event reconstruction procedure
was applied. Finally, the ratio
of reconstructed events to the number of generated
events gave a combined efficiency/acceptance correction in each kinematic bin.

\begin{figure}[!h]
\begin{center}
\includegraphics[bb=1cm 6cm 20cm 23cm, scale=0.4]{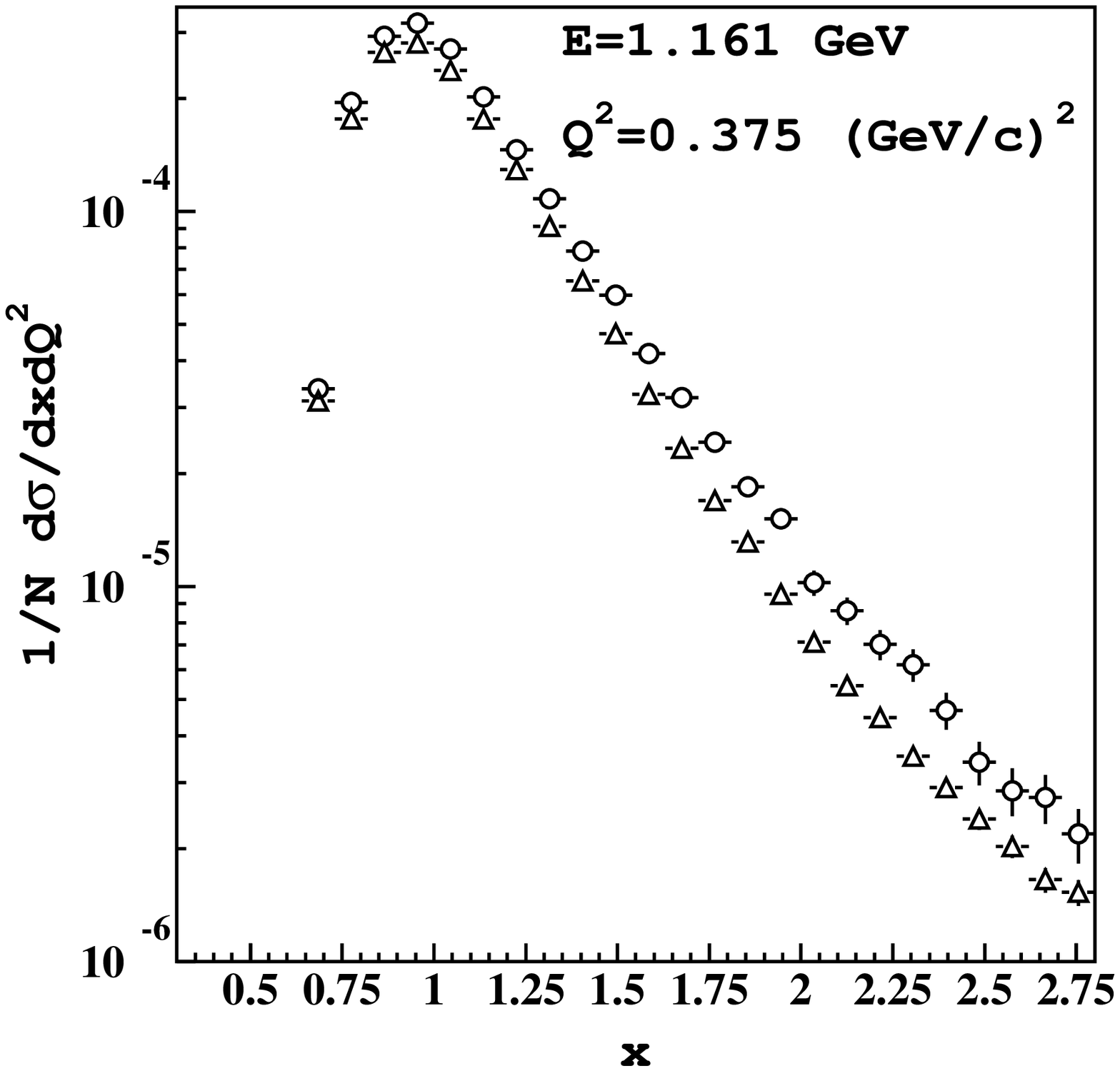}~
\includegraphics[bb=1cm 6cm 20cm 23cm, scale=0.4]{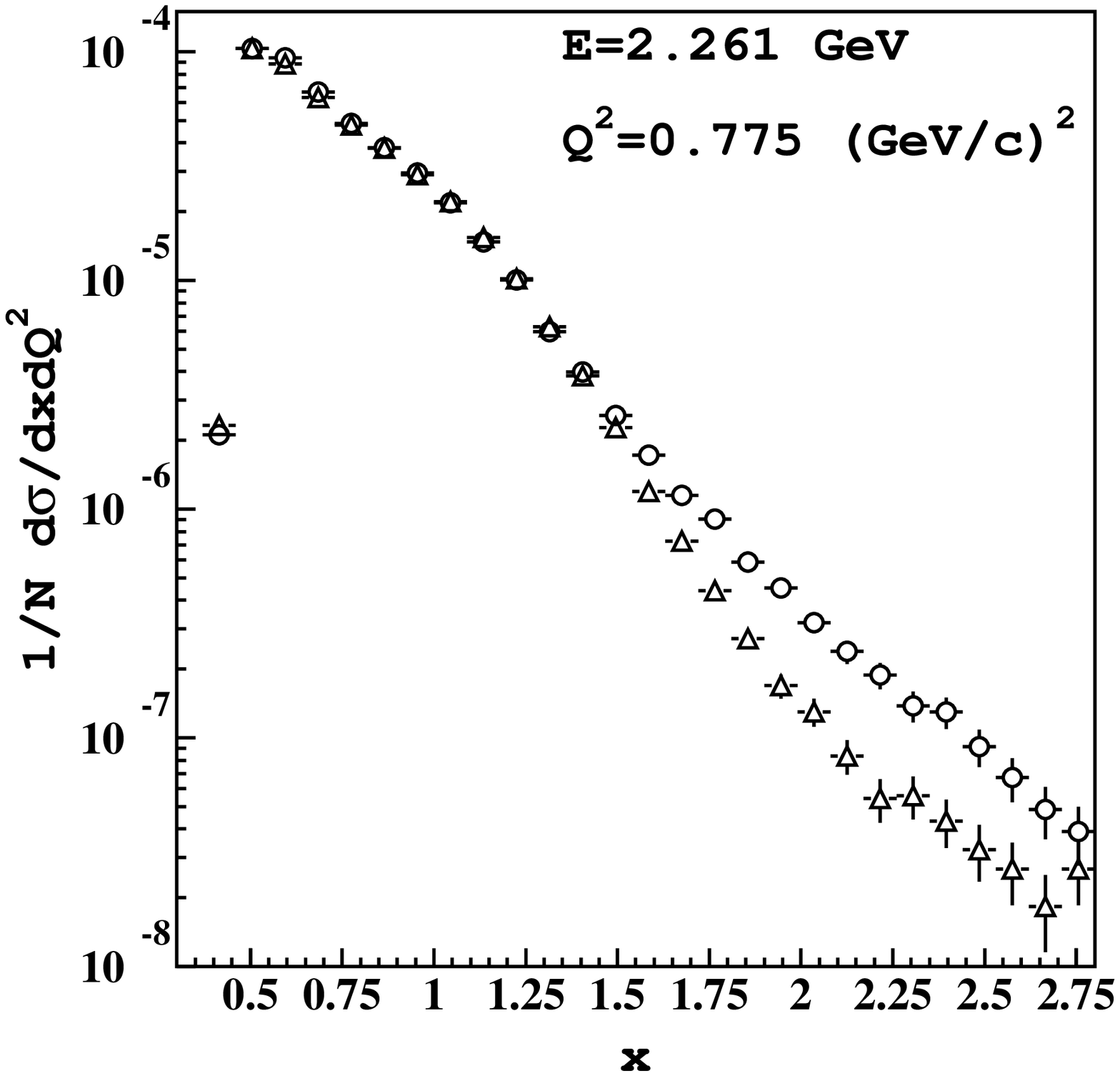}
\includegraphics[bb=1cm 6cm 20cm 23cm, scale=0.4]{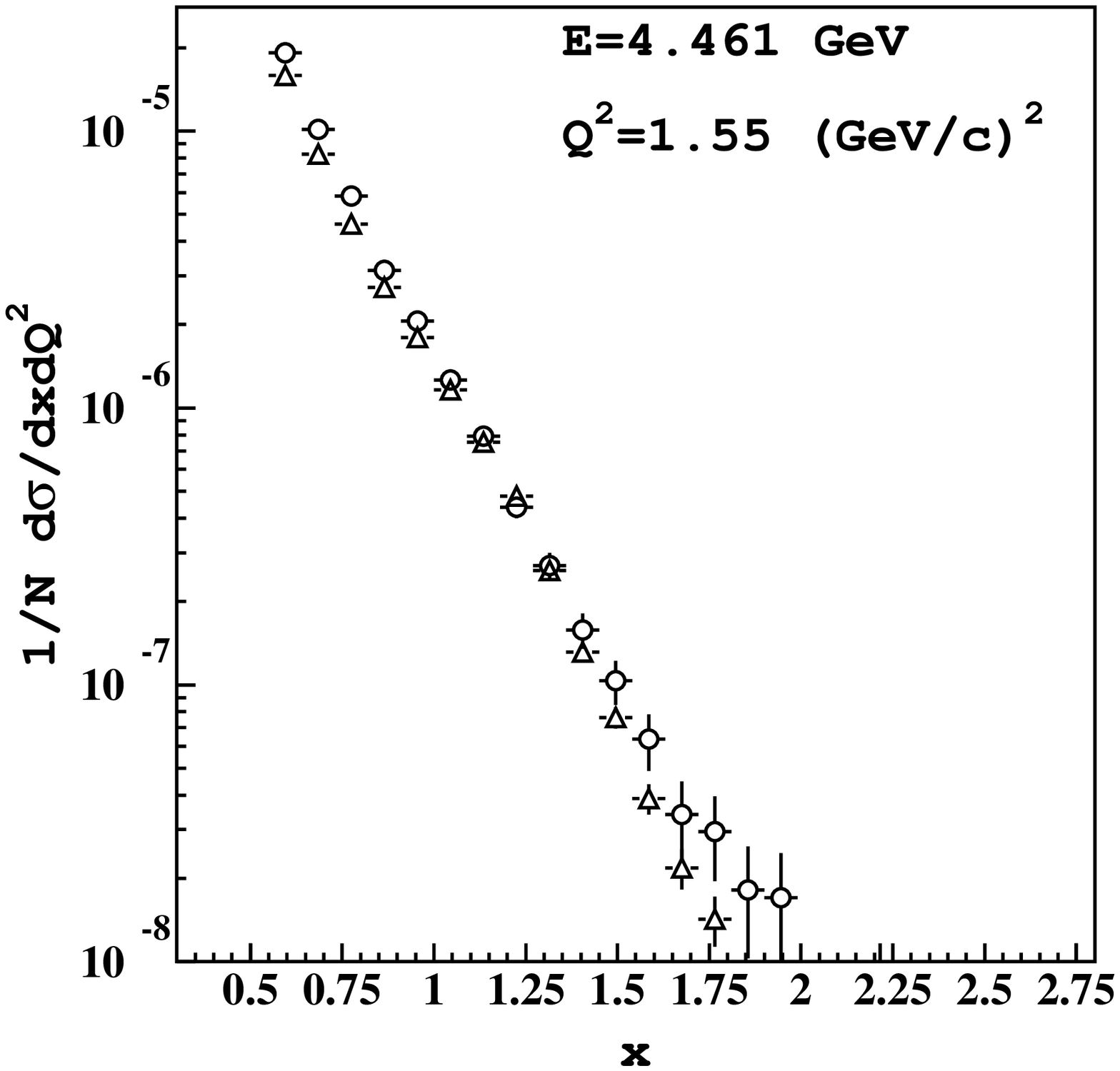}~
\includegraphics[bb=1cm 6cm 20cm 23cm, scale=0.4]{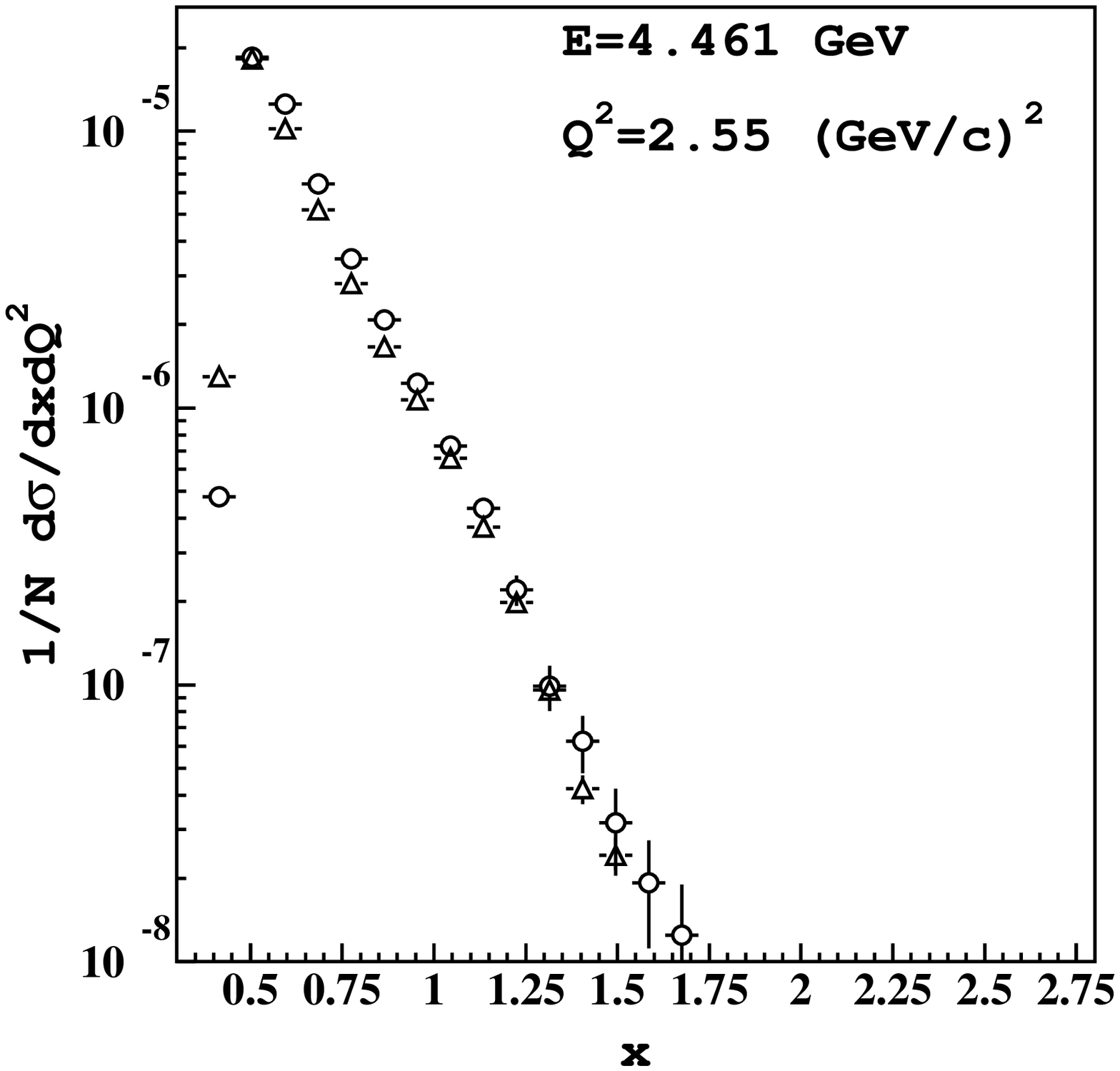}
\caption{\label{fig:mc_data_yields} Comparison of normalized yields from the data (open circles)
and reconstructed Monte Carlo (open triangles) for four different kinematic settings.
The error bars include only statistical uncertainties.}
\end{center}
\end{figure}

Detector response simulations were performed in the same
way as described in Ref.~\cite{osipenko_f2p}.
The following improvements and changes for electron-nucleus
scattering were implemented:
\begin{enumerate}
\item Electron scattering events
were generated by a random event generator with the
probability distributed according to $\sigma^M_{rad}$.
The values for the elastic and inelastic cross sections
for electron-carbon scattering were taken from existing fits of
world data in Refs.~\cite{eC_elastic} and~\cite{Ricco2,Simula_nucl}, respectively.
\item The contribution from internal radiative processes was added according to
calculations~\cite{Mo}.
\item The event rate obtained in the simulations was then compared to the data,
preserving the original normalizations (accumulated FC charge
for the data and the number of generated events over the integrated cross
section of the event generator for simulations). These normalized yields
differ from the cross section by the acceptance, efficiency and radiative corrections.
The simulated events were subjected to the same cuts applied to the data.
$e^+e^-$ pair production
and empty target backgrounds were subtracted from the data.
The normalized yields obtained with the same set of cuts from
the data and simulations were compared and found to be in
good agreement at $x<1.5$ (see Fig.~\ref{fig:mc_data_yields}).
At larger $x$, the contribution of nucleon short range correlations (SRC),
which are not included in the model, may play a significant role~\cite{Egiyan03,Egiyan06}.
\end{enumerate}
\noindent 
The efficiency appears to be fairly flat and not lower than 97\% inside
the region of the detector defined by the fiducial cuts.

\subsection{Structure Function $F_2(x,Q^2)$}\label{sec:d_sf}
The measured electron yield $N_C$, normalized to the integrated luminosity 
in conjunction with Monte Carlo simulations, were used to extract
the structure function $F_2$ in each kinematic bin.
The Monte Carlo events were used to simultaneously obtain efficiency,
acceptance, bin centering and radiative corrections. $F_2$ has been determined using:
\begin{equation}
F_2(x,Q^2) =
\frac{1}{\rho \frac{N_A}{M_A} A L Q_{tot}} 
\frac{J}{\sigma_{Mott}} \frac{\nu}{1+\frac{1-\epsilon}{\epsilon}\frac{1}{1+R}}
\frac{N_C(x,Q^2)}
{\Psi(x,Q^2)}
F_{e^+e^-}(x,Q^2)~,
\label{eq:d_ii2}
\end{equation}
\noindent
where $\nu=E_0-E^\prime$ is the energy transfer,
$\rho$ is the density of the carbon target, $N_A$ is
the Avogadro constant, $M_A$ is the target molar mass, $L$ is the
target length, $Q_{tot}$ is the total charge in the Faraday Cup (FC)
and $\Psi(x,Q^2)$ is the efficiency including the radiative
and bin-centering correction factors:
\begin{equation}
\Psi(x,Q^2)=\Psi_{eff}(x,Q^2)\Psi_{rad}(x,Q^2)
\Psi_{bin}(x,Q^2)~,
\label{eq:d_ii3}
\end{equation}
\noindent
with:
\begin{equation}
\Psi_{rad}=\frac{\sigma^M_{rad}}{\sigma^M}~~~
\mbox{and}~~~\Psi_{bin}
= \frac{\int_{\Delta \tau} d\sigma^M}{\sigma^M}~.
\label{eq:d_ii4}
\end{equation}
\noindent
$\Psi_{eff}$ is the ratio between the number of reconstructed and
generated events in the bin and $\sigma^M$ is the event generator model cross section.
The integral in Eq.~\ref{eq:d_ii4} was taken over the bin area $\Delta \tau$.
Here $\epsilon$ is the virtual photon polarization parameter:
\begin{equation}
\epsilon \equiv \Biggl(
1+2\frac{\nu^2+Q^2}{Q^2} \tan^2{\frac{\theta}{2}}
\Biggr )^{-1}~.
\label{eq:d_sf2}
\end{equation}
\noindent 
The Mott cross section $\sigma_{Mott}$ and the Jacobian $J$
of the transformation between $d\Omega dE^\prime$ to $dx dQ^2$
are defined by:
\begin{equation}
\sigma_{Mott}=\frac{\alpha^2 \cos^2{\frac{\theta}{2}}}{4E_0^2\sin^4{\frac{\theta}{2}}}
~~~\mbox{and}~~~
J=\frac{x E_0 E^{\prime}}{\pi \nu}~.
\end{equation}
\noindent 
The structure function $F_2(x,Q^2)$ was extracted using
the fit of the function $R(x,Q^2)\equiv \sigma_L / \sigma_T$
described in Ref.~\cite{osipenko_f2d}.
However, the structure function
$F_2$ in the relevant kinematic range is very insensitive
to the value of $R$.
For example, at the typical kinematics of
$\epsilon = 0.75$ and assuming a SLAC DIS $R$ value of 0.18, the relative
uncertainties of $F_2$ and $R$ are related by $\Delta F_2/F_2=0.03 \Delta R/R$.
Therefore, at this kinematics a relative error in $R$ of 20\%  will generate
only a relative error of 0.6\% in $F_2$.
The overlapped data from two different beam
energies were combined using a weighted-average technique.
Moreover, we checked that the parametrization used for $R$
was consistent with the difference between the two cross sections
within statistical and systematic uncertainties.

\begin{figure*}
\begin{center}
\includegraphics[bb=1cm 6cm 19cm 23cm, scale=0.4]{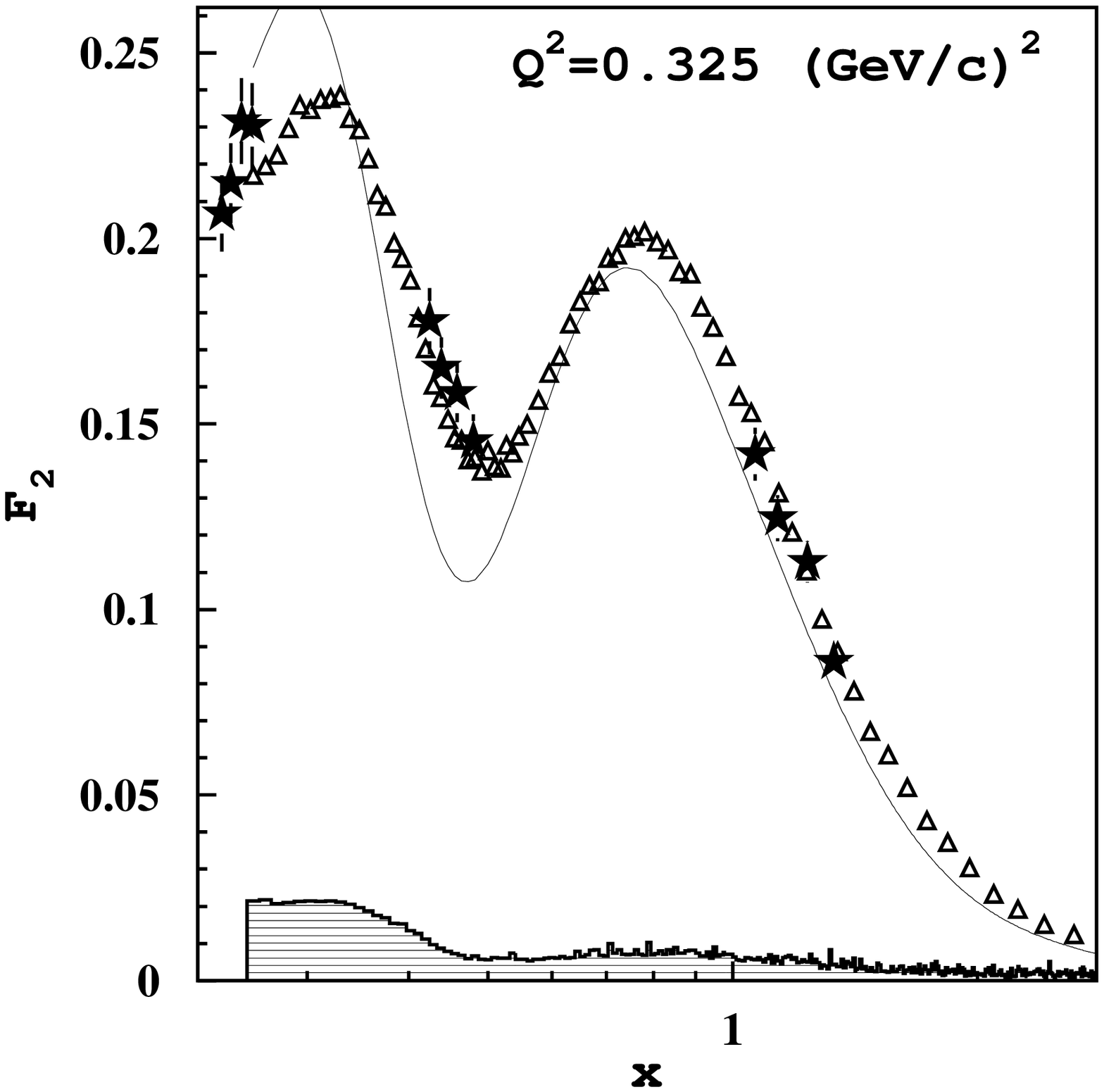}~%
\includegraphics[bb=1cm 6cm 19cm 23cm, scale=0.4]{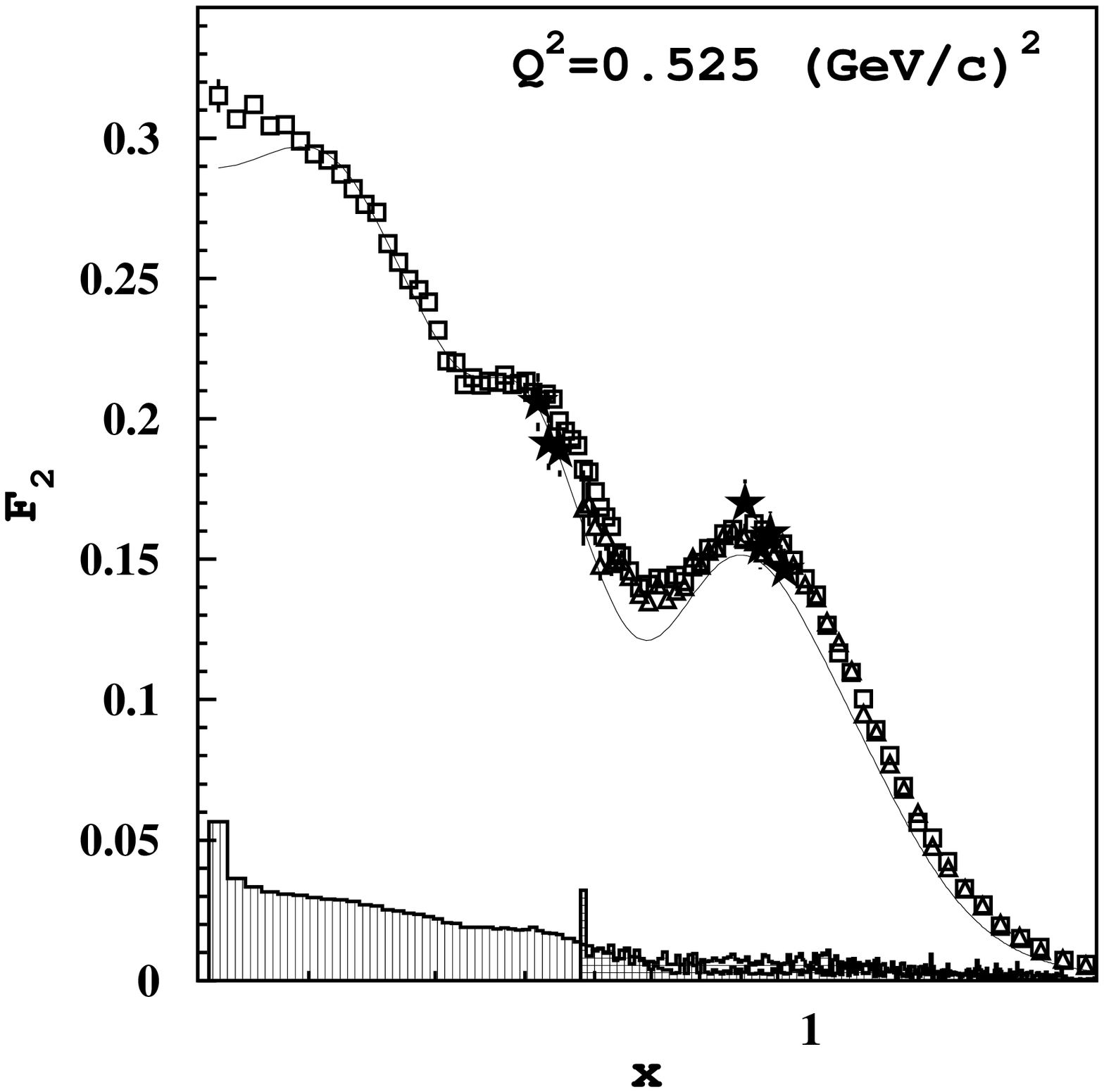}
\includegraphics[bb=1cm 6cm 19cm 23cm, scale=0.4]{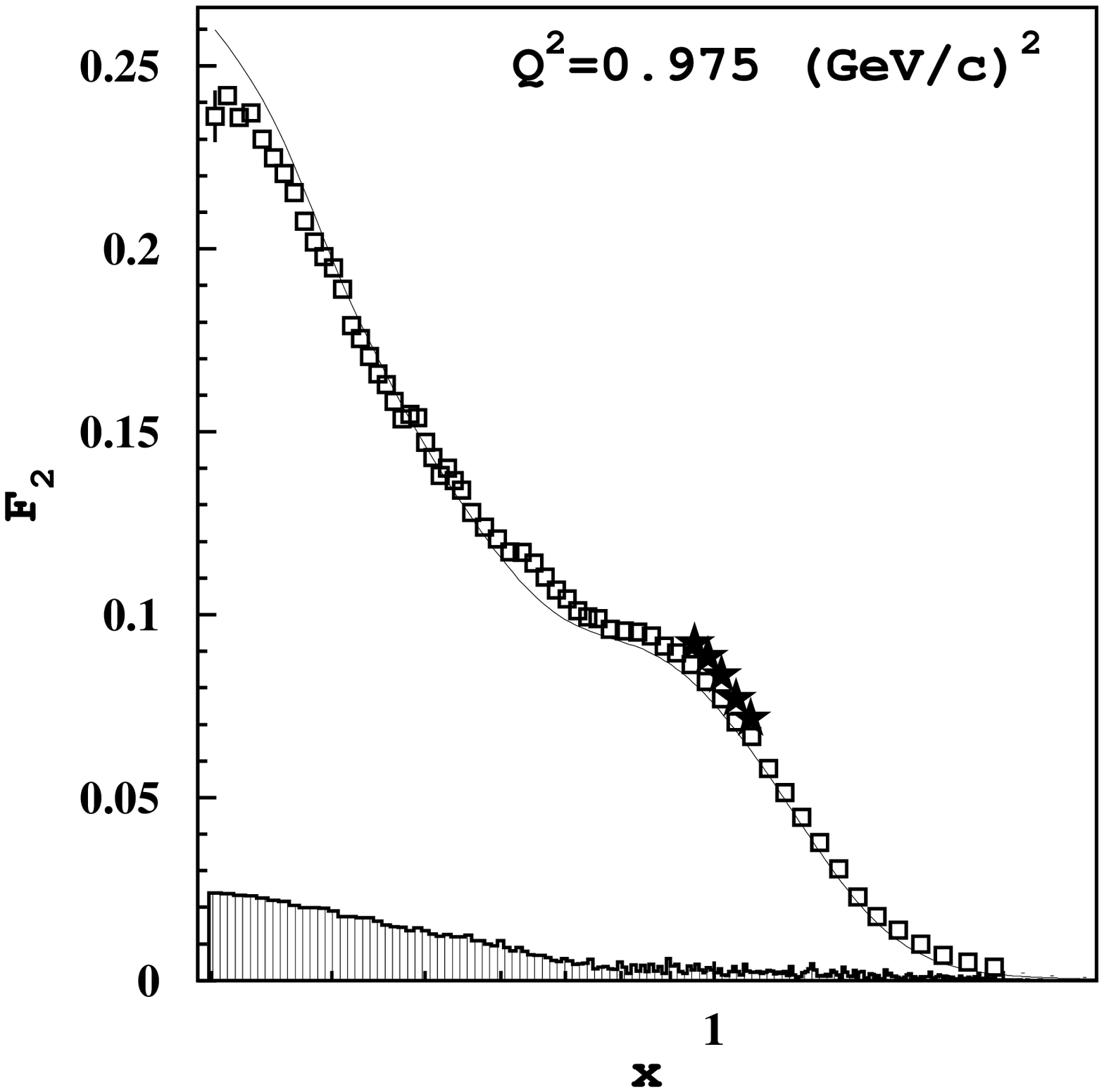}~%
\includegraphics[bb=1cm 6cm 19cm 23cm, scale=0.4]{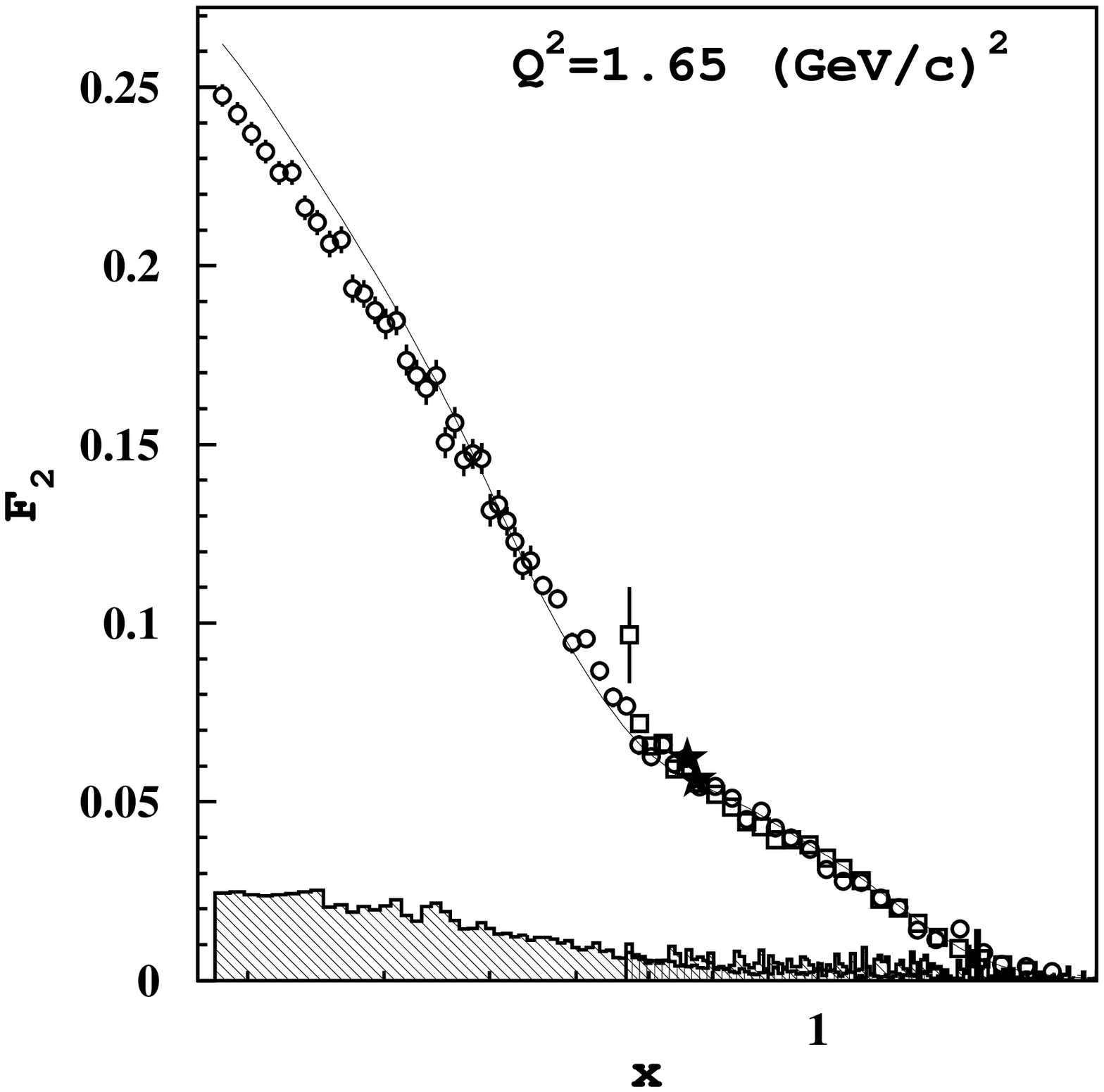}
\includegraphics[bb=1cm 6cm 19cm 23cm, scale=0.4]{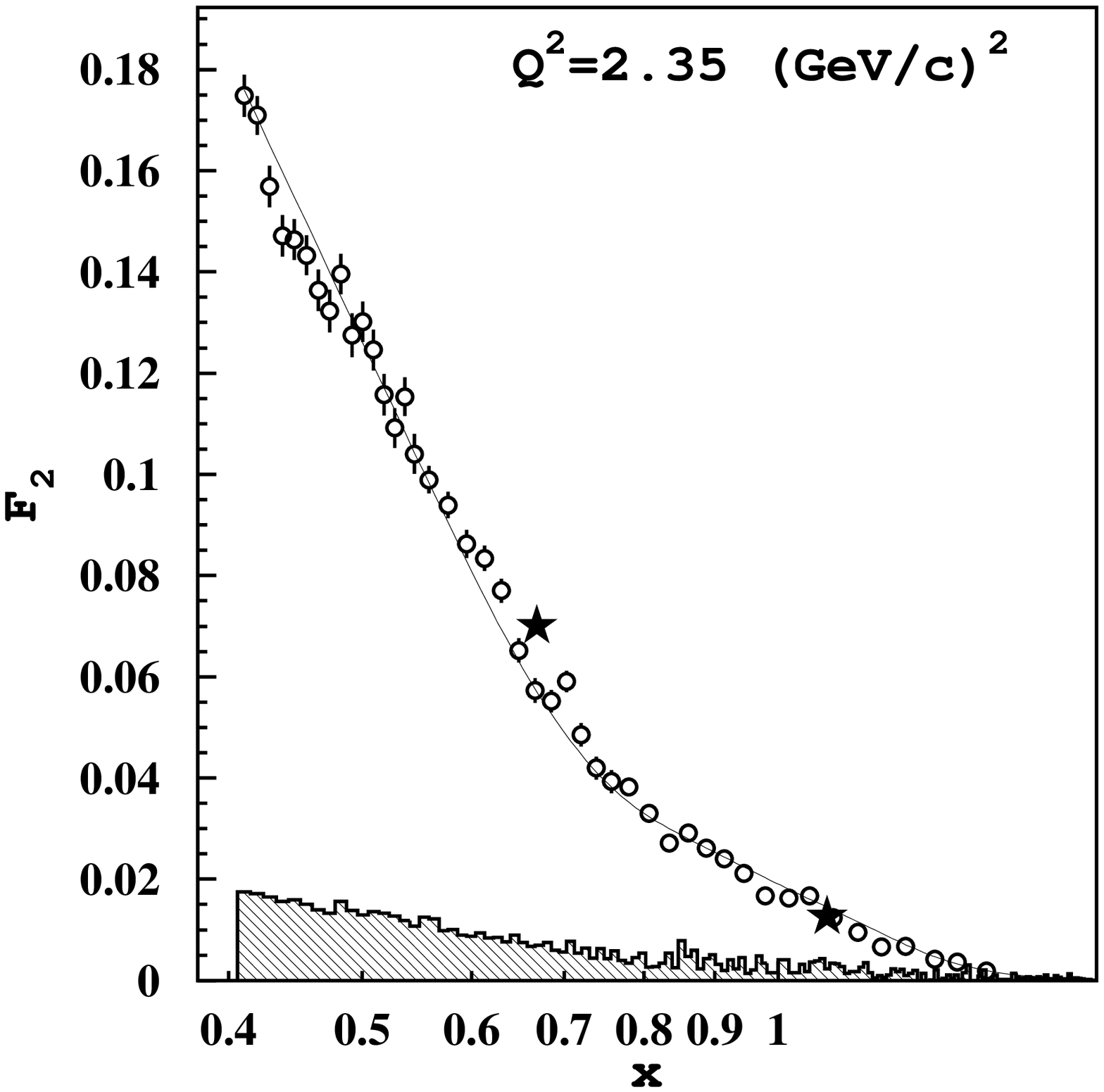}~%
\includegraphics[bb=1cm 6cm 19cm 23cm, scale=0.4]{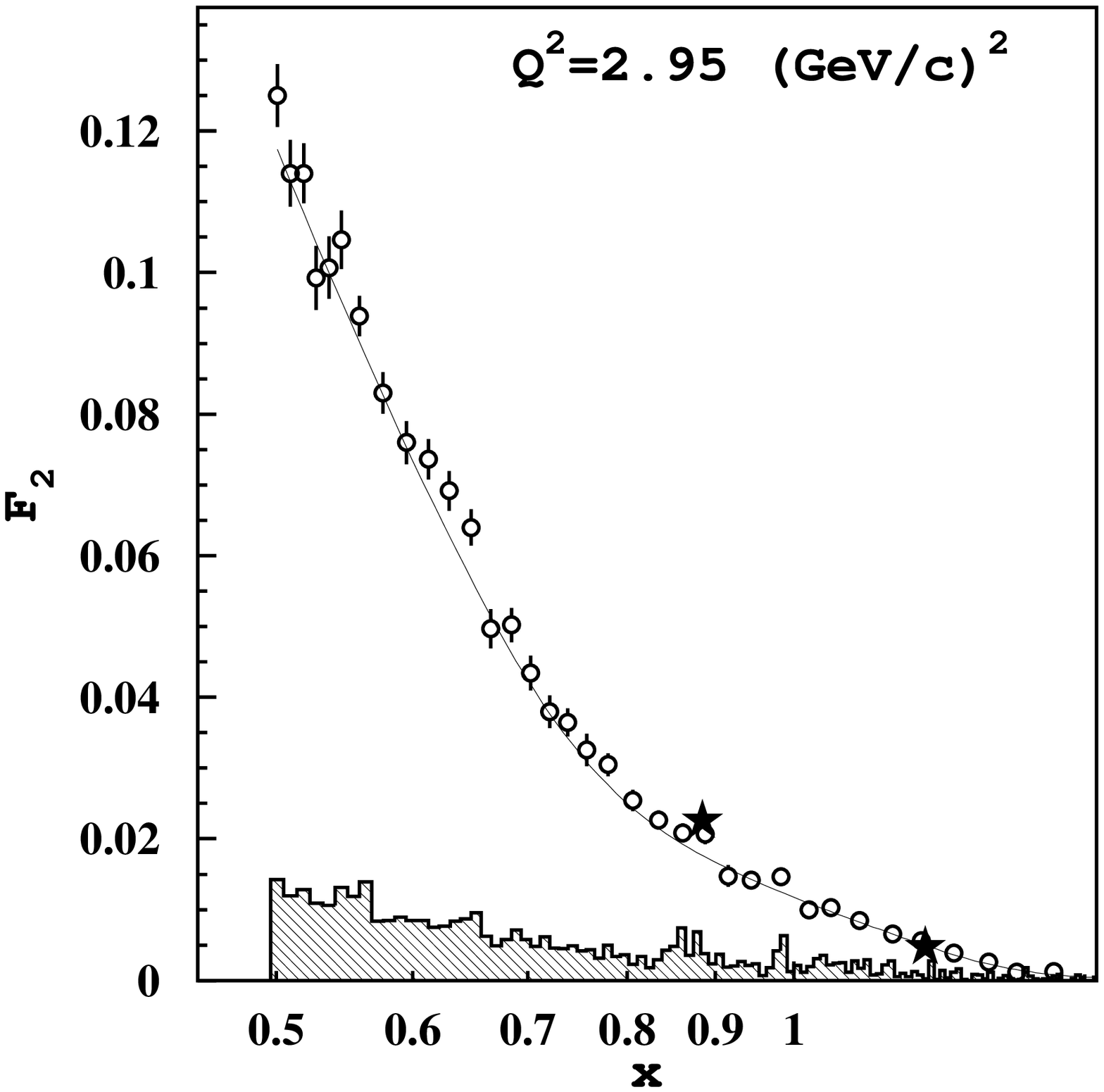}
\caption{\label{fig:f2comp} The carbon structure function $F_2(x,Q^2)$
per nucleon at six different $Q^2$ values.
The open symbols represent experimental data obtained in the present analysis
(triangles for $E_0=1.161$ GeV, squares for $E_0=2.261$ GeV and circles for $E_0=4.461$ GeV)
with systematic uncertainties indicated by the hatched areas.
The full stars show data from previous experiments~\cite{gomez94,baran88,day93,
desy73,hall_c,kiev69,oconnell84,saclay83,sealock89,yerevan88,Harvard71,Arrington95,Whitney74}.
The solid line represents the parametrization from Refs.~\cite{Ricco2,Simula_nucl}.}
\end{center}
\end{figure*}

Fig.~\ref{fig:f2comp} shows a comparison between the $F_2$ data
from CLAS and the other world data in a few $Q^2$ bins.
The CLAS data agree very well with all previous measurements.
The measured $F_2(x,Q^2)$ values are given in the center of
the corresponding $x$ and $Q^2$ bin. The Monte Carlo based bin centering correction
from Eq.~\ref{eq:d_ii4} allows to interpolate the data averaged over the bin
to its value in the middle of the bin. Therefore, we do not
provide any uncertainty in the determination of $x$ and $Q^2$.
However, this uncertainty is implicitly included in the systematic uncertainties
of Monte Carlo simulations. Indeed, Monte Carlo simulations include
the effect of event migration from the original bin to neighbor bins,
due to the finite detector resolution and energy loss in the target.
Hence, the loss or gain of events in each single bin in the data is compensated
by a similar change of the number of Monte Carlo events.
As a test, we verified that the data obtained within significantly larger,
not rectangular bins (in which bin migration is suppressed)
agree with the presented data within statistical uncertainties.
The values of $F_2(x,Q^2)$, together with their statistical and systematic
uncertainties, are tabulated elsewhere~\cite{f2a_database}.

The radiative correction factor $\Psi_{rad}$ was calculated in the following way:
\begin{itemize}
\item the $eC$ elastic radiative tail was calculated according to the ``exact''
Mo and Tsai formula~\cite{Mo};
\item in the quasi-elastic peak region ($W^{el}+\Delta W<W<1.2$ GeV) we have applied
the correction formula to the continuum spectrum given in Ref.~\cite{Mo},
which is
based on the peaking approximation and is known to be reliable only
when $E^\prime/E_0 > 0.5$. Here $W^{el}$ is the $eC$ elastic peak position
and $\Delta W$ is its width;
\item at $W>1.2$ GeV we applied the exact Mo and Tsai formula to the
quasi-elastic tail, while a formula based on the
peaking approximation (referred to as the ``unfolding procedure'')
was used for the inelastic spectrum. For an exact calculation of the
quasi-elastic tail,
it was necessary to extract quasi-elastic form-factors. To this end
we integrated the quasi-elastic cross section given by our parametrization from the beginning
of the peak up to $W=1.2$ GeV and performed a separation of the electric and magnetic
form-factors. The tails calculated in these two kinematic regions
do not exhibit any discontinuity at the point $W=1.2$ GeV. This
assured us that the peaking approximation formalism is safely applicable
to the quasi-elastic tail up to $W=1.2$ GeV.
\end{itemize}
The radiative correction factor $\Psi_{rad}$ varies strongly
in the explored kinematic range from 0.7 up to 1.5. Fortunately,
the largest corrections are given by the tails of the elastic and
quasi-elastic peaks,
for which calculations are very accurate (of the order of percent, see Refs.~\cite{Mo,Akushevich}).

\subsection{Systematic Uncertainties}\label{sec:d_su}
Here we summarize all the systematic uncertainties taken into account in the
present analysis:
\begin{itemize}
\item The cross section overall normalization carries a systematic
uncertainty due to the known precision of the target length and thickness.
We estimated this uncertainty to be 1\%. For the lowest beam energy set,
$E_0=1.161$ GeV, an additional systematic uncertainty
due to FC charge loss corrections of 3\% was added.

\item Previous CLAS inclusive measurements~\cite{osipenko_f2p,osipenko_f2d} on proton and deuteron targets
showed that the combined efficiency/acceptance systematic uncertainty of CLAS data
averaged to 4.5\%. We assumed a 3\% uniform systematic uncertainty
due to the efficiency evaluation, while the systematic uncertainty on the acceptance,
based on Monte Carlo simulations, was evaluated separately.

\item There are two systematic uncertainties in the Monte Carlo simulation.
The first one is due to the model dependence of the
reaction cross section used for generating the events.
Taking into account the good agreement between the data and Monte Carlo
simulations we neglected this uncertainty.
The second systematic uncertainty is due to
the inability of the GEANT3-based CLAS
simulation package GSIM~\cite{GSIM} to reproduce the CLAS response
to electron tracks at different
angles and momenta. To estimate this effect, we treated the six CLAS sectors
as independent spectrometers. The normalized event yield measured
in each sector was compared separately to the simulations,
as shown in Fig.~\ref{fig:mc_data_yields}.
The observed differences were compared sector-by-sector to remove
uncertainties due to the event generator model.
From this comparison, we obtained an average systematic
uncertainty varying for different data sets from 5 to 8\%.

\item The $e^+e^-$-pair production correction introduces a systematic 
uncertainty due to the parametrization involved in the calculations.
We assumed 30\% uncertainty on the inclusive $e^+$ cross section~\cite{BostedPP}
and 10\% on the inelastic electron-carbon cross section~\cite{Ricco2,Simula_nucl}.
The systematic uncertainty is kinematics dependent, but since the
correction is very close to unity, the systematic uncertainty
associated with it never exceeds 5\%.

\item $e^-$-identification is based on the CC signal.
The correlation of the CC to a particle track is dependent on the
magnitude of the CC signal. Therefore, the systematic
uncertainty is related to the mean number of photoelectrons 
in each kinematic point. The inefficiency of the CC 
signal is given by $[1-\exp{(-N_{phe})}]$,
where $N_{phe}$ is the mean number of photoelectrons.
We took this probability as an estimate of the systematic
uncertainty due to $e^-$-identification.

\item The systematic uncertainties of the radiative corrections (RC)
were estimated by changing the various parametrizations used in 
the RC evaluation, and computing the differences with the standard 
parametrization. In particular, we changed the elastic $^{12}$C 
form-factor by 10\%, the quasi-elastic cross section 
by $[10+40/(1+Q^2/M^2)]$ \% and the inelastic cross section by 10\%. The 
three obtained differences were combined in quadrature to 
extract the final systematic uncertainty of the radiative 
corrections.

\item Since we applied corrections to the measured $e^-$ momentum, we 
estimated the related systematic uncertainty. The momentum 
correction that brings the hydrogen elastic peak to the 
correct position results in an average momentum shift of
0.003$\Delta p/p$
for $E_0=1.161$ GeV and of 0.0035$\Delta p/p$ for $E_0=2.261$ GeV.
For $E_0=4.461$ GeV we assumed that the possible momentum
shift can be slightly larger than the two measured values
for the lower energies, 0.004$\Delta p/p$. We shifted the entire
data spectra by the abovementioned amounts to evaluate the relative
change of the cross section in each bin.
This modified relative change gave us an estimate of the uncertainty
on the measured cross section.
For the two lower
beam energies, these relative shifts were suppressed by the
factor accounting for the fixed value of the momentum
at the hydrogen elastic peak position. This factor, which reduced
the uncertainty in the vicinity of
the hydrogen elastic peak, was parametrized as
$[1-\exp{(-(\nu-Q^2/2M)^2/(2\sigma_p^2))}]$,
where the width parameter $\sigma_p$ was taken to be 0.3 GeV for the $E_0=1.161$ GeV data
and 0.4 GeV for the $E_0=2.261$ GeV data.
For the $E_0=4.461$ GeV data set, the relative change in
the cross section was taken directly as an estimate of the momentum
measurement uncertainty.

\item Thanks to recent measurements of $R=\frac{\sigma_L}{\sigma_T}$
performed at Jefferson Lab~\cite{rlt_hallc},
the precision of this quantity is greatly improved in our kinematic domain.
Moreover, the structure function
$F_2$ in the interesting kinematic range is not very sensitive
to the value of $R$. In fact, even a 100\% systematic uncertainty
on $R$ gives only a few percent uncertainty on $F_2$.
\begin{equation}\label{eq:d_sf6}
\frac{\Delta F_2}{F_2}(x,Q^2)=
\frac{1-\epsilon}{1+\epsilon R}
\frac{\Delta R}{1+R} \approx R (1-\epsilon)\frac{\Delta R}{R} ~~.
\end{equation}
\noindent 
Uncertainties of $R$ given in Ref.~\cite{osipenko_f2d} were propagated
to $F_2$ and the corresponding systematic uncertainties
were always lower than 5\%.
\end{itemize}

All systematic uncertainties mentioned above and listed in Table \ref{table:r_ps}
were combined in quadrature to give the total systematic uncertainty.
When data sets from different beam energies overlapped, the systematic
uncertainties were calculated as the central values using a weighted-average technique.

\begin{table}[!h]
\begin{center}
\caption{Average systematic uncertainties for the three data sets
for each of the different sources.}
\label{table:r_ps}
\vspace{2mm}
\begin{tabular}{|c|c|c|c|} \hline
                                             & \multicolumn{3}{|c|}{Data set} \\ \cline{2-4}
Source of uncertainty                        & $E_0=1.161$ GeV & $E_0=2.261$ GeV & $E_0=4.461$ GeV \\
                                             &  $[\%]$ &  $[\%]$ &  $[\%]$  \\ \hline
Normalization                                &   3.2   &    1	 &    1     \\ \hline
Efficiency                                   &   3     &    3	 &    3     \\ \hline
Acceptance                                   &   6.6   &   4.7   &   7.7    \\ \hline
$e^+e^-$-pair production                     &   0     &   0.001 &   0.05   \\ \hline
$e^-$ identification                         &   0.04  &   0.4   &   0.2    \\ \hline
Radiative correction                         &   1.3   &   1.7   &    5     \\ \hline
Momentum measurement                         &   0.9   &   0.03  &   12     \\ \hline
Uncertainty on $R=\frac{\sigma_L}{\sigma_T}$ &   2     &   1.2   &   0.5    \\ \hline
Total                                        &   8.5   &   6.5   &   20     \\ \hline
\end{tabular}
\end{center}
\end{table}

\subsection{Moments of the Structure Function $F_2$}\label{sec:d_nm}
The non-zero mass of the target leads to an undesirable mixing between QCD operators
of different spin in the OPE.
To avoid this mixing in our analysis, we use Nachtmann~\cite{Nachtmann} moments $M^{NM}_n(Q^2)$
instead of the usual (massless) Cornwall-Norton moments.
The Nachtmann moments are defined as follows:
\begin{equation}
M^{NM}_n(Q^2) =\int_0^A dx \frac{\xi^{n+1}}{x^3} F_2(x,Q^2)
\Biggl[\frac{3+3(n+1)r+n(n+2)r^2}{(n+2)(n+3)}\Biggr] ,
\label{eq:i_nm1}
\end{equation}
\noindent
where $r = \sqrt{1+4M^2x^2/Q^2}$, $M$ is the nucleon mass and $\xi = 2x/(1+r)$.
In the Bjorken limit $M^2/Q^2 \to 0$, the 
Nachtmann and Cornwall-Norton definitions coincide.

\begin{figure}
\begin{center}
\includegraphics[bb=1cm 6cm 20cm 23cm, scale=0.4]{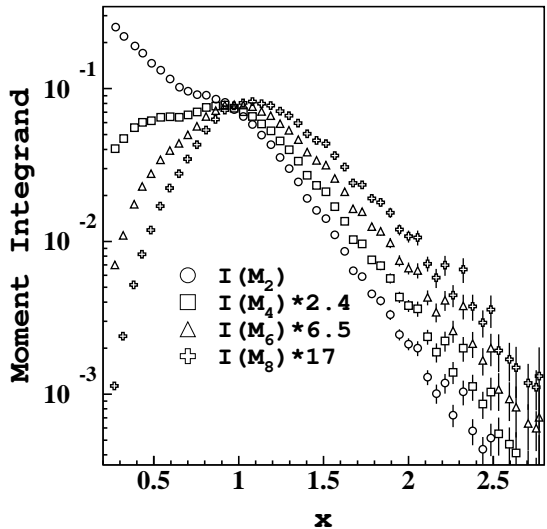}
\caption{\label{fig:intgr} Integrands of the Nachtmann moments ($I(M_n)$) at
$Q^2=0.825$ (GeV/c)$^2$: circles represent the integrand of $M_2$;
squares show the integrand of $M_4$;
triangles show the integrand of $M_6$;
crosses show the integrand of $M_8$.}
\end{center}
\end{figure}

The carbon structure function moments were evaluated 
according to the method described in Refs.~\cite{osipenko_f2p,osipenko_f2d}.
The CLAS data were combined with
the remaining world data on the structure function $F_2$, along with the inclusive
cross section data
from Refs.~\cite{gomez94,baran88,day93,desy73,hall_c,kiev69,oconnell84,saclay83,
sealock89,yerevan88,Harvard71,Arrington95,Whitney74,
bcdms94,bcdms,cern81,emc,cornell76,daresbury79,desy77,fermilab81,e665,emc88_p1,emc88_p2,emc89,nmc95}
(see Fig.~\ref{fig:xandQ2Domain}).
The $Q^2$-range of the CLAS data, from 0.175 to 4.95 (GeV/c)$^2$,
was divided
into bins of width $\Delta Q^2 =$ 0.05-0.1~(GeV/c)$^2$. Within each
$Q^2$ bin, the world data were
shifted to the central bin value $Q^2_0$, using the fit of $F_2(x,Q^2)$
from Refs.~\cite{Ricco2,Simula_nucl}.
The integrals of the data over $x$ were performed numerically using the
standard trapezoidal method TRAPER~\cite{cernlib}.
As an example, Fig.~\ref{fig:intgr} shows the integrands of the first four
moments
as a function of $x$ at fixed $Q^2$. The significance of the large $x$
region for various moments can clearly be seen.

\begin{figure}
\begin{center}
\includegraphics[bb=1cm 6cm 20cm 23cm, scale=0.6]{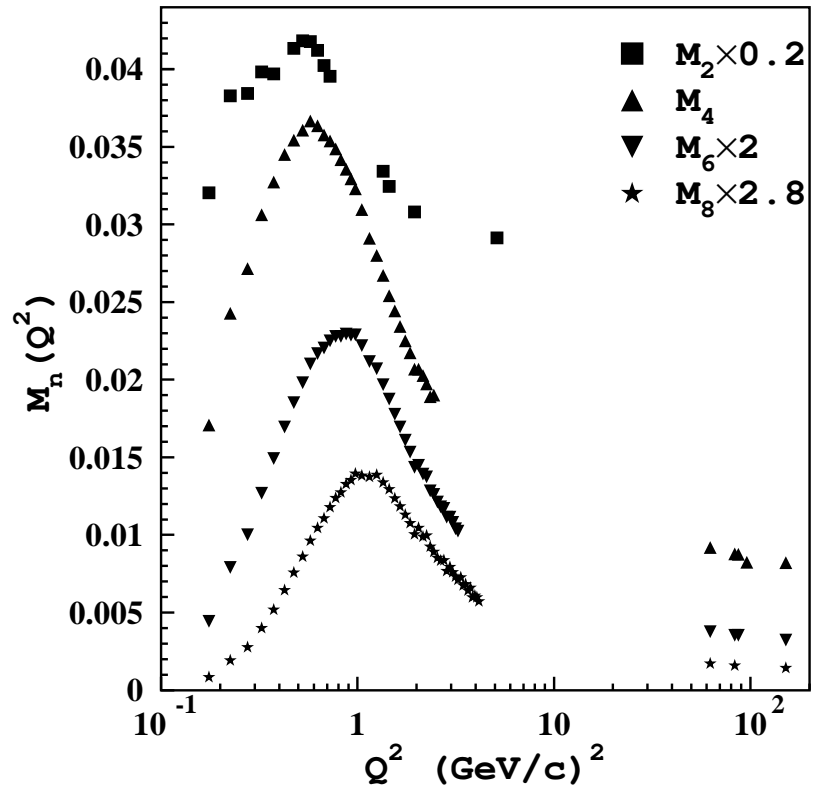}
\caption{\label{fig:NachtMom}
The Nachtmann moments extracted from the world data,
including the new CLAS results. Uncertainties are statistical only
and since their values do not exceed a few percent (see Table \ref{table:r_nm1}),
they are smaller than the symbol size.}
\end{center}
\end{figure}

As in Refs.~\cite{osipenko_f2p,osipenko_f2d}, the world data at $Q^2$ above
$5$~(GeV/c)$^2$ were analyzed in the same way as described above,
but with a different $Q^2$ bin size.
The bin size was chosen to provide sufficient $x$-coverage for most of the
$Q^2$ bins ($\Delta Q^2/Q^2 =5$ \%).
The results, together with their statistical and
systematic uncertainties, are shown in Fig.~\ref{fig:NachtMom}
and reported in Table~\ref{table:r_nm1}.
A comparison of Fig.~\ref{fig:NachtMom} to the corresponding
figures in Refs.~\cite{osipenko_f2p,osipenko_f2d} shows
the lack of the carbon data in the $Q^2$ interval from $5$ to $40$ (GeV/c)$^2$.
This kinematic interval is very important because here moments
reach the scaling regime.

\begin{figure}
\begin{center}
\includegraphics[bb=1cm 6cm 20cm 23cm, scale=0.4]{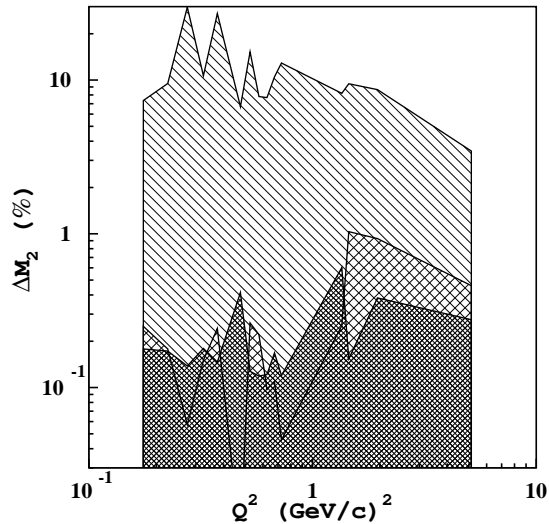}
\caption{\label{fig:SepErr} Uncertainties of the Nachtmann moment $M_2$
in percent. The lower cross-hatched area represents the statistical uncertainties.
The left-hatched area represents the systematic uncertainties. The right-hatched
area represents the low-$x$ extrapolation uncertainty.}
\end{center}
\end{figure}

The systematic uncertainty consists (see for example, Fig.~\ref{fig:SepErr})
of experimental uncertainties in our data and
the data given in Refs.~\cite{gomez94,baran88,day93,desy73,hall_c,kiev69,oconnell84,saclay83,sealock89,
yerevan88,Harvard71,Arrington95,Whitney74,
bcdms94,bcdms,cern81,emc,cornell76,daresbury79,desy77,fermilab81,e665,emc88_p1,emc88_p2,emc89,nmc95},
and uncertainties in the evaluation procedure. 
The first type of uncertainties was accounted for in the moment evaluation point-by-point.
To estimate them we had to account for the inclusion of many data sets
measured in different laboratories with different detectors.
In the present analysis, we assumed that the
different experiments are independent and, therefore,
only the systematic uncertainties within a given data set are correlated.
The uncertainties in the evaluation procedure were obtained
as described previously in Refs.~\cite{osipenko_f2p,osipenko_f2d}.
The lowest moment, $n=2$, includes a systematic uncertainty
due to the low-$x$ extrapolation, which is negligible for larger $n$.
To estimate this uncertainty we modified the parametrization from Refs.~\cite{Ricco2,Simula_nucl},
used for the low-$x$ extrapolation.
The modification, evaluated by a fit to the available carbon $F_2$ data in the low-$x$ region,
was given by the multiplicative factor: $1+(0.38+0.044/Q^2) x^{0.3} (1-x)^{1.75}$.
At $Q^2=1$ (GeV/c)$^2$ the relative change in the value of $F_2$ parametrization
ranges from 2\% at $x\sim 10^{-5}$ up to 18\% at $x\sim 10^{-1}$.
Then we compared the moments from the Table~\ref{table:r_nm1}
with ones obtained by using the modified low-$x$ extrapolation
and we took the difference as the estimate of the systematic uncertainty.
As one can see from Fig.~\ref{fig:SepErr}, systematic
uncertainties dominate at $n=2$, while the statistical
uncertainties become comparable at larger $n$.

\begin{sidewaystable}
\caption{\label{table:r_nm1}The Nachtmann moments for
$n=2,4,6$ and $8$ evaluated in the interval
$0.15 \le~Q^2 \le 150$~(GeV/c)$^2$. The moments are labeled
with an asterisk when the contribution to the integral by the experimental
data is
between 50\% and 70\%. All the others were evaluated
with more than 70\% data coverage. The data are reported
together with the statistical and systematic uncertainties,
the third uncertainty for $n=2$ is due to the low-$x$
extrapolation.}
\begin{tabular}{|c|c|c|c|c|} \cline{1-5}
$Q^2~[$(GeV/c)$^2]$ & $M_2(Q^2)$x$10^{-1}$ & $M_4(Q^2)$x$10^{-2}$ & $M_6(Q^2)$x$10^{-2}$ & $M_8(Q^2)$x$10^{-3}$ \\ \cline{1-5}
  0.175 & 1.602 $\pm$ 0.003 $\pm$ 0.126 $\pm$ 0.004 & 1.707 $\pm$ 0.004 $\pm$ 0.133 & 0.222 $\pm$ 0.001 $\pm$ 0.018 & 0.302 $\pm$ 0.001 $\pm$ 0.027 \\ \cline{1-5}
  0.225 & 1.915 $\pm$ 0.003 $\pm$ 0.223 $\pm$ 0.003 & 2.428 $\pm$ 0.007 $\pm$ 0.179 & 0.395 $\pm$ 0.002 $\pm$ 0.029 & 0.687 $\pm$ 0.003 $\pm$ 0.051 \\ \cline{1-5}
  0.275 & 1.923 $\pm$ 0.003 $\pm$ 0.645 $\pm$ 0.001 & 2.715 $\pm$ 0.005 $\pm$ 0.303 & 0.500 $\pm$ 0.001 $\pm$ 0.045 & 0.990 $\pm$ 0.002 $\pm$ 0.092 \\ \cline{1-5}
  0.325 & 1.991 $\pm$ 0.004 $\pm$ 0.256 $\pm$ 0.003 & 3.063 $\pm$ 0.006 $\pm$ 0.198 & 0.635 $\pm$ 0.001 $\pm$ 0.040 & 1.424 $\pm$ 0.004 $\pm$ 0.100 \\ \cline{1-5}
  0.375 & 1.985 $\pm$ 0.003 $\pm$ 0.592 $\pm$ 0.005 & 3.273 $\pm$ 0.007 $\pm$ 0.271 & 0.746 $\pm$ 0.002 $\pm$ 0.044 & 1.854 $\pm$ 0.006 $\pm$ 0.111 \\ \cline{1-5}
  0.425 &                                           & 3.452 $\pm$ 0.008 $\pm$ 0.215 & 0.847 $\pm$ 0.002 $\pm$ 0.057 & 2.298 $\pm$ 0.008 $\pm$ 0.180 \\ \cline{1-5}
  0.475 & 2.067 $\pm$ 0.009 $\pm$ 0.145 $\pm$ 0.000 & 3.544 $\pm$ 0.009 $\pm$ 0.219 & 0.926 $\pm$ 0.003 $\pm$ 0.059 & 2.707 $\pm$ 0.011 $\pm$ 0.188 \\ \cline{1-5}
  0.525 & 2.091 $\pm$ 0.003 $\pm$ 0.391 $\pm$ 0.005 & 3.608 $\pm$ 0.008 $\pm$ 0.231 & 0.990 $\pm$ 0.003 $\pm$ 0.062 & 3.073 $\pm$ 0.011 $\pm$ 0.222 \\ \cline{1-5}
  0.575 & 2.090 $\pm$ 0.002 $\pm$ 0.176 $\pm$ 0.005 & 3.667 $\pm$ 0.008 $\pm$ 0.194 & 1.051 $\pm$ 0.003 $\pm$ 0.057 & 3.435 $\pm$ 0.013 $\pm$ 0.216 \\ \cline{1-5}
  0.625 & 2.061 $\pm$ 0.002 $\pm$ 0.150 $\pm$ 0.002 & 3.636 $\pm$ 0.009 $\pm$ 0.209 & 1.085 $\pm$ 0.004 $\pm$ 0.065 & 3.730 $\pm$ 0.017 $\pm$ 0.260 \\ \cline{1-5}
  0.675 & 2.011 $\pm$ 0.003 $\pm$ 0.254 $\pm$ 0.002 & 3.577 $\pm$ 0.010 $\pm$ 0.225 & 1.102 $\pm$ 0.005 $\pm$ 0.067 & 3.949 $\pm$ 0.024 $\pm$ 0.276 \\ \cline{1-5}
  0.725 & 1.977 $\pm$ 0.002 $\pm$ 0.326 $\pm$ 0.001 & 3.537 $\pm$ 0.010 $\pm$ 0.242 & 1.125 $\pm$ 0.005 $\pm$ 0.072 & 4.208 $\pm$ 0.028 $\pm$ 0.313 \\ \cline{1-5}
  0.775 &                                           & 3.488 $\pm$ 0.011 $\pm$ 0.213 & 1.139 $\pm$ 0.006 $\pm$ 0.069 & 4.417 $\pm$ 0.034 $\pm$ 0.324 \\ \cline{1-5}
  0.825 &                                           & 3.418 $\pm$ 0.009 $\pm$ 0.205 & 1.139 $\pm$ 0.004 $\pm$ 0.072 & 4.545 $\pm$ 0.024 $\pm$ 0.332 \\ \cline{1-5}
  0.875 &                                           & 3.357 $\pm$ 0.008 $\pm$ 0.201 & 1.148 $\pm$ 0.005 $\pm$ 0.074 & 4.741 $\pm$ 0.027 $\pm$ 0.369 \\ \cline{1-5}
  0.925 &                                           & 3.295 $\pm$ 0.008 $\pm$ 0.327 & 1.143 $\pm$ 0.005 $\pm$ 0.082 & 4.835 $\pm$ 0.027 $\pm$ 0.405 \\ \cline{1-5}
  0.975 &                                           & 3.230 $\pm$ 0.009 $\pm$ 0.229 & 1.144 $\pm$ 0.005 $\pm$ 0.066 & 4.976 $\pm$ 0.030 $\pm$ 0.289 \\ \cline{1-5}
\end{tabular}
\end{sidewaystable}
\begin{sidewaystable}
\begin{tabular}{|c|c|c|c|c|} \cline{1-5}
$Q^2~[$(GeV/c)$^2]$ & $M_2(Q^2)$x$10^{-1}$ & $M_4(Q^2)$x$10^{-2}$ & $M_6(Q^2)$x$10^{-2}$ & $M_8(Q^2)$x$10^{-3}$ \\ \cline{1-5}
  1.050 &                                           & 3.097 $\pm$ 0.006 $\pm$ 0.205 & 1.109 $\pm$ 0.004 $\pm$ 0.065 & 4.936 $\pm$ 0.023 $\pm$ 0.277 \\ \cline{1-5}
  1.150 &                                           & 2.911 $\pm$ 0.007 $\pm$ 0.234 & 1.058 $\pm$ 0.004 $\pm$ 0.077 & 4.911 $\pm$ 0.029 $\pm$ 0.402 \\ \cline{1-5}
  1.250 &                                           & 2.799 $\pm$ 0.007 $\pm$ 0.259 & 1.035 $\pm$ 0.005 $\pm$ 0.081 & 4.957 $\pm$ 0.035 $\pm$ 0.429 \\ \cline{1-5}
  1.350 & 1.672 $\pm$ 0.010 $\pm$ 0.130 $\pm$ 0.004 & 2.673 $\pm$ 0.007 $\pm$ 0.221 & 0.984 $\pm$ 0.004 $\pm$ 0.080 & 4.778 $\pm$ 0.034 $\pm$ 0.421 \\ \cline{1-5}
  1.450 &*1.622 $\pm$ 0.002 $\pm$ 0.141 $\pm$ 0.017 & 2.541 $\pm$ 0.008 $\pm$ 0.215 & 0.938 $\pm$ 0.005 $\pm$ 0.076 & 4.627 $\pm$ 0.034 $\pm$ 0.385 \\ \cline{1-5}
  1.550 &                                           & 2.443 $\pm$ 0.008 $\pm$ 0.254 & 0.889 $\pm$ 0.005 $\pm$ 0.081 & 4.408 $\pm$ 0.039 $\pm$ 0.416 \\ \cline{1-5}
  1.650 &                                           & 2.343 $\pm$ 0.008 $\pm$ 0.267 & 0.847 $\pm$ 0.005 $\pm$ 0.083 & 4.232 $\pm$ 0.036 $\pm$ 0.437 \\ \cline{1-5}
  1.750 &                                           & 2.250 $\pm$ 0.009 $\pm$ 0.238 & 0.805 $\pm$ 0.005 $\pm$ 0.089 & 4.033 $\pm$ 0.038 $\pm$ 0.489 \\ \cline{1-5}
  1.850 &                                           & 2.174 $\pm$ 0.010 $\pm$ 0.244 & 0.766 $\pm$ 0.005 $\pm$ 0.096 & 3.844 $\pm$ 0.040 $\pm$ 0.575 \\ \cline{1-5}
  1.950 &*1.540 $\pm$ 0.006 $\pm$ 0.114 $\pm$ 0.014 & 2.067 $\pm$ 0.010 $\pm$ 0.221 & 0.718 $\pm$ 0.005 $\pm$ 0.085 & 3.582 $\pm$ 0.036 $\pm$ 0.473 \\ \cline{1-5}
  2.050 &                                           & 2.068 $\pm$ 0.012 $\pm$ 0.235 & 0.725 $\pm$ 0.007 $\pm$ 0.096 & 3.731 $\pm$ 0.047 $\pm$ 0.570 \\ \cline{1-5}
  2.150 &                                           & 2.027 $\pm$ 0.012 $\pm$ 0.257 & 0.696 $\pm$ 0.007 $\pm$ 0.098 & 3.533 $\pm$ 0.050 $\pm$ 0.586 \\ \cline{1-5}
  2.250 &                                           & 1.974 $\pm$ 0.013 $\pm$ 0.228 & 0.687 $\pm$ 0.008 $\pm$ 0.105 & 3.560 $\pm$ 0.066 $\pm$ 0.663 \\ \cline{1-5}
  2.350 &                                           & 1.891 $\pm$ 0.013 $\pm$ 0.225 & 0.642 $\pm$ 0.009 $\pm$ 0.113 & 3.300 $\pm$ 0.080 $\pm$ 0.791 \\ \cline{1-5}
  2.450 &                                           & 1.901 $\pm$ 0.013 $\pm$ 0.424 & 0.630 $\pm$ 0.009 $\pm$ 0.130 & 3.180 $\pm$ 0.087 $\pm$ 0.814 \\ \cline{1-5}
  2.550 &                                           &                               & 0.605 $\pm$ 0.008 $\pm$ 0.096 & 3.042 $\pm$ 0.065 $\pm$ 0.614 \\ \cline{1-5}
  2.650 &                                           &                               & 0.590 $\pm$ 0.009 $\pm$ 0.104 & 2.983 $\pm$ 0.082 $\pm$ 0.721 \\ \cline{1-5}
  2.750 &                                           &                               & 0.585 $\pm$ 0.009 $\pm$ 0.092 & 2.986 $\pm$ 0.075 $\pm$ 0.615 \\ \cline{1-5}
  2.850 &                                           &                               & 0.557 $\pm$ 0.008 $\pm$ 0.104 & 2.741 $\pm$ 0.069 $\pm$ 0.643 \\ \cline{1-5}
  2.950 &                                           &                               & 0.557 $\pm$ 0.009 $\pm$ 0.094 & 2.831 $\pm$ 0.082 $\pm$ 0.603 \\ \cline{1-5}
\end{tabular}
\end{sidewaystable}
\begin{sidewaystable}
\begin{tabular}{|c|c|c|c|c|} \cline{1-5}
$Q^2~[$(GeV/c)$^2]$ & $M_2(Q^2)$x$10^{-1}$ & $M_4(Q^2)$x$10^{-2}$ & $M_6(Q^2)$x$10^{-2}$ & $M_8(Q^2)$x$10^{-3}$ \\ \cline{1-5}
  3.050 &                                           &                               & 0.540 $\pm$ 0.009 $\pm$ 0.095 & 2.717 $\pm$ 0.075 $\pm$ 0.611 \\ \cline{1-5}
  3.150 &                                           &                               & 0.520 $\pm$ 0.008 $\pm$ 0.092 & 2.616 $\pm$ 0.075 $\pm$ 0.601 \\ \cline{1-5}
  3.250 &                                           &                               & 0.513 $\pm$ 0.008 $\pm$ 0.103 & 2.536 $\pm$ 0.070 $\pm$ 0.642 \\ \cline{1-5}
  3.350 &                                           &                               &                               & 2.597 $\pm$ 0.068 $\pm$ 0.655 \\ \cline{1-5}
  3.450 &                                           &                               &                               & 2.409 $\pm$ 0.069 $\pm$ 0.579 \\ \cline{1-5}
  3.550 &                                           &                               &                               & 2.437 $\pm$ 0.056 $\pm$ 0.747 \\ \cline{1-5}
  3.650 &                                           &                               &                               & 2.295 $\pm$ 0.078 $\pm$ 0.765 \\ \cline{1-5}
  3.750 &                                           &                               &                               & 2.354 $\pm$ 0.072 $\pm$ 0.613 \\ \cline{1-5}
  3.850 &                                           &                               &                               & 2.137 $\pm$ 0.063 $\pm$ 0.589 \\ \cline{1-5}
  3.950 &                                           &                               &                               & 2.170 $\pm$ 0.064 $\pm$ 0.599 \\ \cline{1-5}
  4.050 &                                           &                               &                               & 2.132 $\pm$ 0.072 $\pm$ 0.557 \\ \cline{1-5}
  4.150 &                                           &                               &                               & 2.038 $\pm$ 0.071 $\pm$ 0.692 \\ \cline{1-5}
  5.125 &*1.457 $\pm$ 0.004 $\pm$ 0.047 $\pm$ 0.007 &                               &                               &                               \\ \cline{1-5}
 62.500 &                                           & 0.920 $\pm$ 0.007 $\pm$ 0.023 & 0.189 $\pm$ 0.002 $\pm$ 0.002 & 0.608 $\pm$ 0.009 $\pm$ 0.009 \\ \cline{1-5}
 83.250 &                                           & 0.877 $\pm$ 0.009 $\pm$ 0.016 & 0.177 $\pm$ 0.003 $\pm$ 0.001 & 0.561 $\pm$ 0.010 $\pm$ 0.007 \\ \cline{1-5}
 87.000 &                                           &*0.874 $\pm$ 0.007 $\pm$ 0.026 & 0.176 $\pm$ 0.002 $\pm$ 0.007 &                               \\ \cline{1-5}
 96.000 &                                           &*0.824 $\pm$ 0.011 $\pm$ 0.030 &                               &                               \\ \cline{1-5}
151.500 &                                           &*0.821 $\pm$ 0.015 $\pm$ 0.018 & 0.162 $\pm$ 0.005 $\pm$ 0.001 & 0.512 $\pm$ 0.020 $\pm$ 0.003 \\ \cline{1-5}
\end{tabular}
\end{sidewaystable}

\section{Phenomenological Twists Expansion}\label{sec:Discussion}
The $Q^2$-distributions of the $n=2,4,6$ and $8$ moments
were fit using the method developed in Refs.~\cite{osipenko_f2p,SIM00,Ricco1}
with the following parametrization:
\begin{equation}
M_n(\alpha_S,Q^2) = \eta_n(\alpha_S) + HT_n(\alpha_S,Q^2) ~ ,
\label{eq:twists}
\end{equation}
\noindent 
where for $n \geq 4$ the leading-twist term contained only one unknown parameter
due to the dominant non-singlet contribution, while for $n=2$,
both singlet and non-singlet terms were considered
(for details see Refs.~\cite{osipenko_f2p,osipenko_f2d}).

The higher-twist contribution is given by~\cite{Ji}:
\begin{equation}
HT_n(\alpha_S,Q^2)= \frac{\mu^2}{Q^2}
\sum_{\tau=4}^{4+2N}
A_{n\tau}\biggl\{\frac{\alpha_s(Q^2)}{\alpha_s(\mu^2)}\biggr\}^{
\gamma_{n \tau}}\biggl[\frac{\mu^2}{Q^2}\biggr]^{\frac{1}{2}(\tau-4)} ~ ,
\label{eq:ht_fit}
\end{equation}
\noindent adding $2N$ free parameters $A_{n\tau}$ and $\gamma_{n\tau}$,
with $N$ being the number of higher-twist terms considered.
Previous analyses~\cite{osipenko_f2p,osipenko_f2d,Ricco1} showed 
that it was necessary to take $N\geq 2$ when $n\geq 4$,
while for $n=2$ it was sufficient to have $N\geq 1$.
In principle, the number of higher-twist terms $N$ should be infinite,
but as was shown in Ref.~\cite{osipenko_confinement06}, already
the minimum number of terms (i.e. $N=1$ for $n=2$ and $N=2$ when $n\geq 4$)
completely described the total higher-twist contribution.
Adding further terms does not change the obtained separation between leading
and higher twists. Hence, in the present analysis we used $N=1$ for $n=2$ and $N=2$ when $n\geq 4$.

The higher-twist parameters $\gamma_{n\tau}$ are effectively the LO anomalous dimensions
of the perturbative Wilson coefficients $E_{n\tau}(\mu_r,\mu_f,\alpha_S)$ in Eq.~\ref{eq:i_m1},
while the parameters $A_{n\tau}$ are proportional to the
matrix elements of QCD operators $O_{n\tau}(\mu)$.
Higher-twist operators get mixed
and only leading-twist parameters $A_{n 2}$ can be directly
related to the operators $O_{n 2}(\mu)$ for $\tau > 2$ since pQCD calculations of $\gamma_{n\tau}$
are at present unavailable.

Based on the above, the $n$-th moment (see Eq.~\ref{eq:ht_fit})
for $n \geq 4$ has five unknown parameters:
the leading-twist parameter $A_{n 2}$ and the higher-twist parameters
$A_{n 4}$, $\gamma_{n 4}$, $A_{n 6}$ and $\gamma_{n 6}$.
These parameters were simultaneously determined from
a $\chi^2$-minimization procedure (MINUIT~\cite{MINUIT})
over the allowed $Q^2$ range.
The statistical uncertainties of the experimental moments were used by MINOS~\cite{MINUIT}
to obtain statistical uncertainties on the extracted parameters.
Their systematic uncertainties were obtained by adding/subtracting
the systematic uncertainties to the experimental moments
and by repeating the twist extraction procedure. 

\begin{figure*}
\begin{center}
\includegraphics[bb=1cm 4cm 20cm 23cm, scale=0.4]{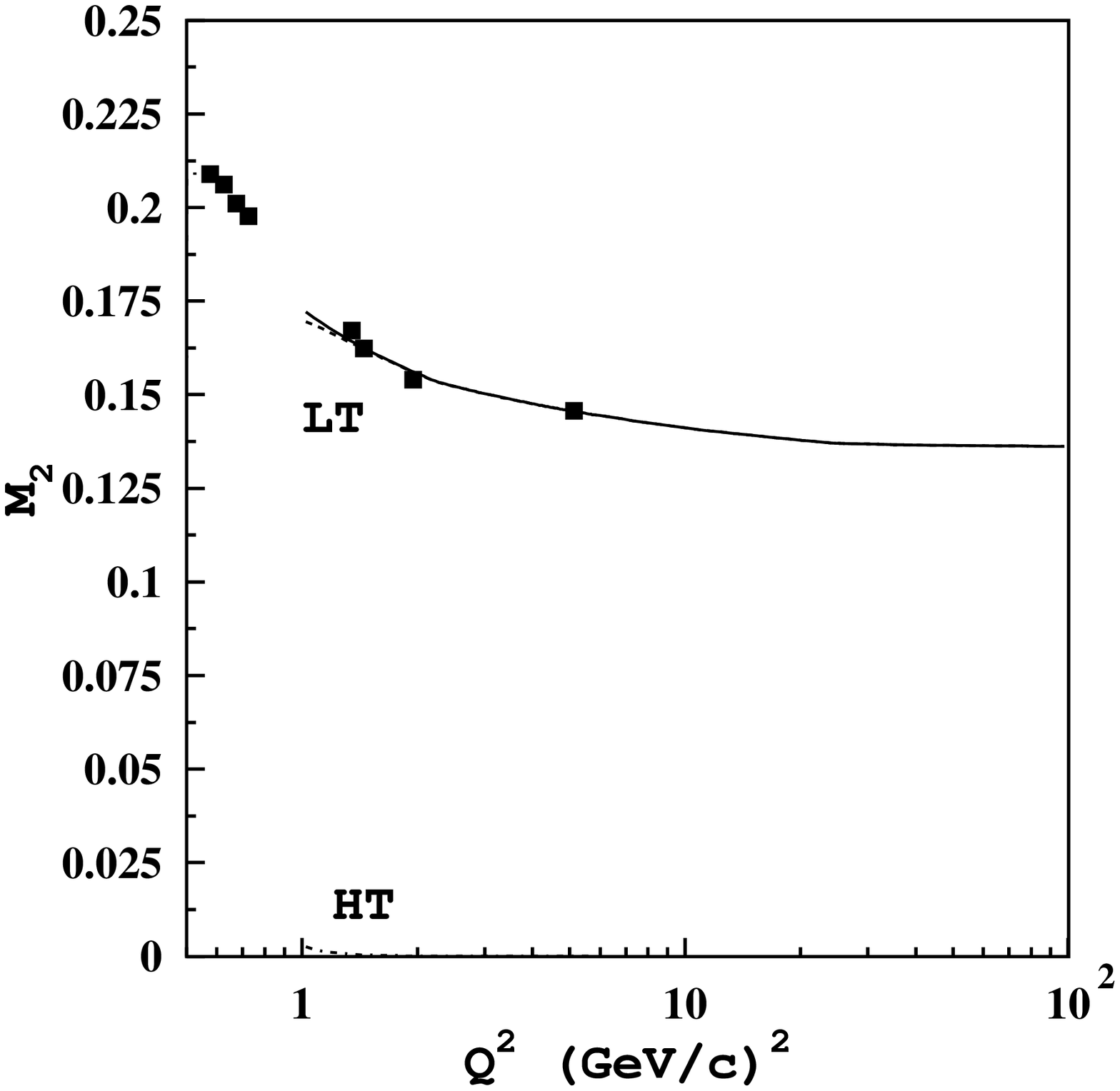}~%
\includegraphics[bb=1cm 4cm 20cm 23cm, scale=0.4]{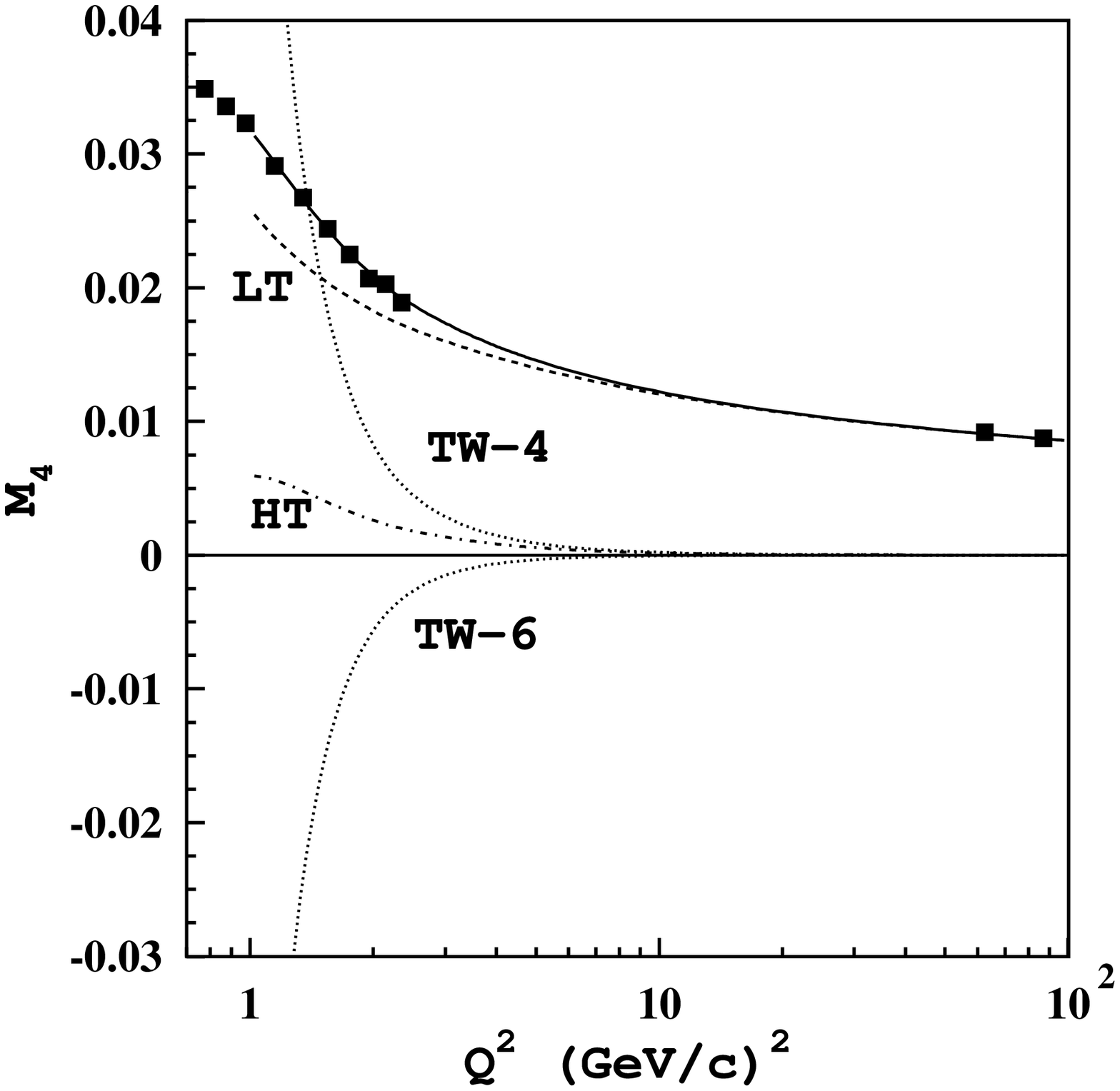}
\includegraphics[bb=1cm 4cm 20cm 23cm, scale=0.4]{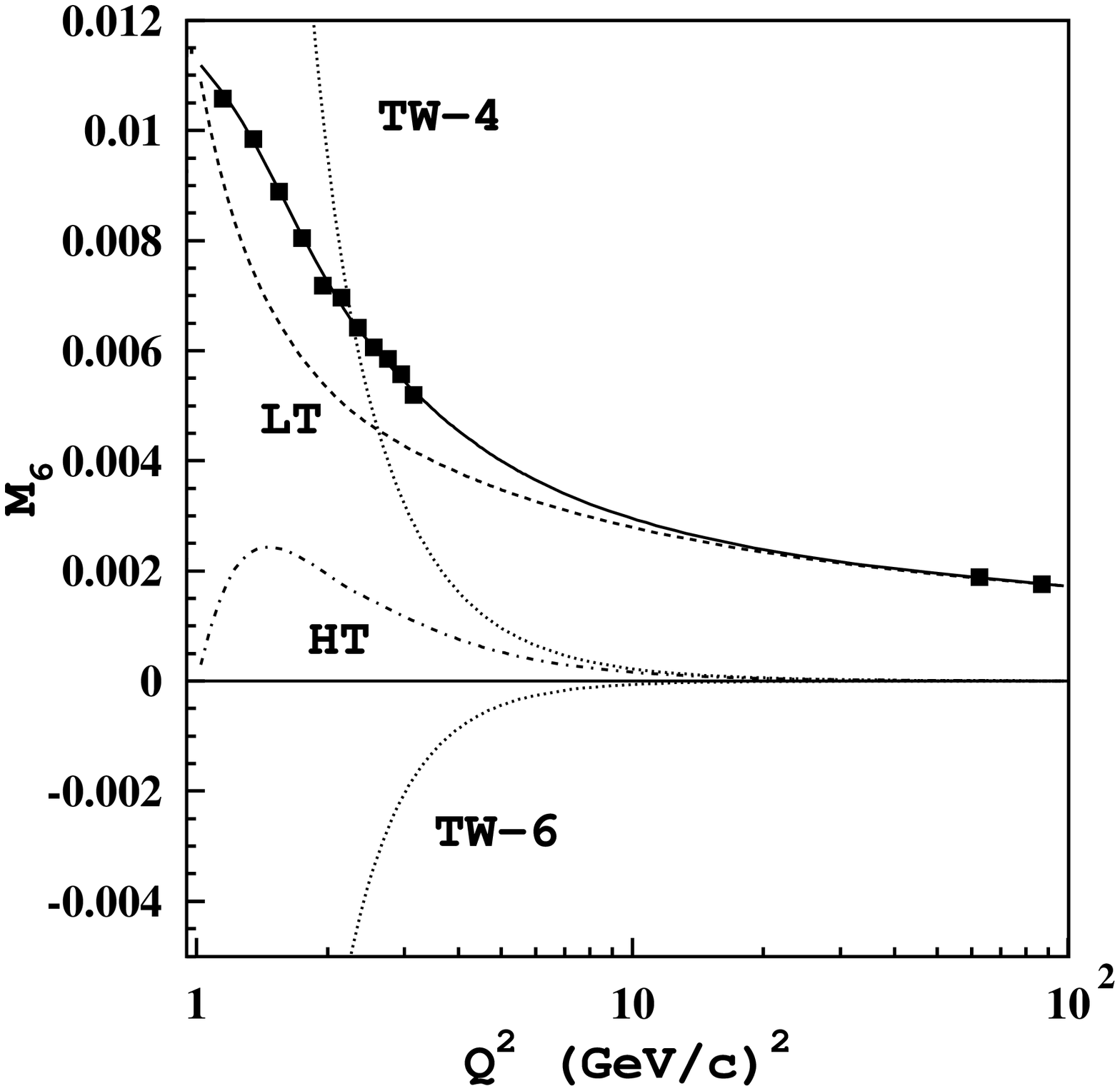}~%
\includegraphics[bb=1cm 4cm 20cm 23cm, scale=0.4]{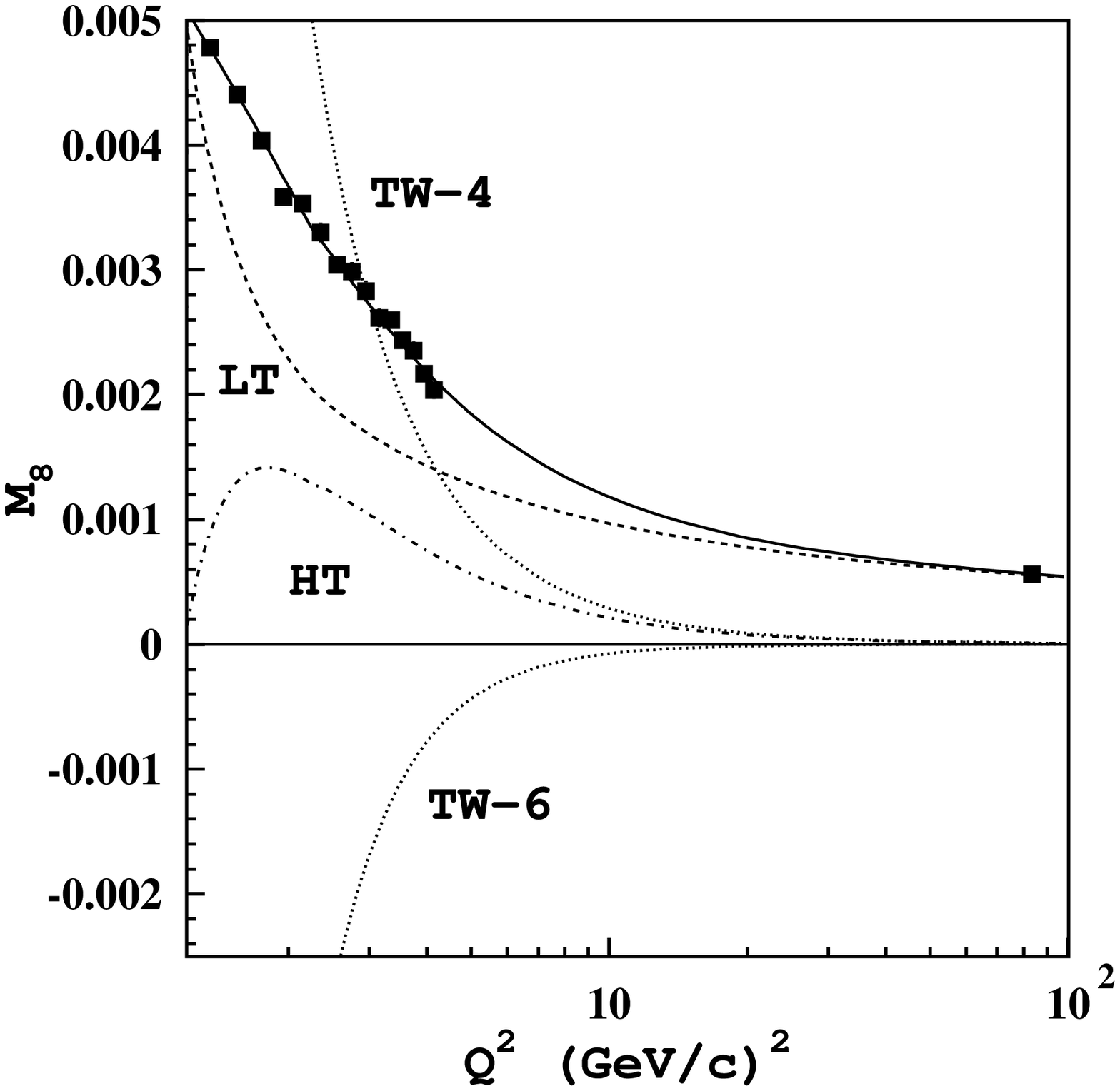}
\caption{\label{fig:twists}
Results of the twist analysis. The squares
represent the Nachtmann moments obtained in this analysis. The
solid line is the fit to the moments using Eq.~\ref{eq:twists} with the
parameters listed in Table~\ref{table:twist1}. The leading-twist (LT), twist-4 (TW-4), twist-6 (TW-6)
and the total higher twist (HT) contributions to the fit are indicated.}
\end{center}
\end{figure*}

In the case of $n=2$, the sum of the non-singlet and
singlet terms at the leading twist was considered,
therefore adding the leading-twist parameter due to the gluon moment at the reference scale.
Moreover, due to the vanishing contribution of the higher twists in $M_2$
(see Fig.~\ref{fig:twists}), we considered the twist-4 term only.

The fit results are shown in Fig.~\ref{fig:twists},
while in Table~\ref{table:twist1}, we report the parameter values
obtained at the reference scale $\mu^2=10$ (GeV/c)$^2$.
In addition, the extracted leading-twist contribution is reported in
Table~\ref{table:ltw}.

\begin{table*}
\caption{\label{table:twist1}Extracted parameters of the twist
expansion at the reference scale $\mu^2=10$ (GeV/c)$^2$.
The first uncertainty has a statistical origin and is obtained from a
MINOS~\cite{cernlib} minimization procedure, while
the second uncertainty is the systematic one described in text.
The contribution of twist-6 to $M_2$ was
too small to be extracted by the present procedure.}
\begin{tabular}{|c|c|c|c|c|} \hline
     n          & 2                                     & 4                                        & 6                                       & 8                                      \\ \hline
$\eta_n(\mu^2)$ &\scriptsize 0.1411$\pm$0.0005$\pm$0.0050         &\scriptsize (1.206$\pm$0.008$\pm$0.051)$\times 10^{-2}$&\scriptsize (2.79$\pm$0.05$\pm$0.07)$\times 10^{-3}$&\scriptsize (9.7$\pm$0.5$\pm$0.3)$\times 10^{-4}$ \\ \hline
$A_{n 4}$       &\scriptsize (0.6$\pm$1.8$\pm$1.1)$\times 10^{-6}$&\scriptsize (2.14$\pm$0.02$\pm$2.68)$\times 10^{-4}$   &\scriptsize (2.22$\pm$0.03$\pm$3.25)$\times 10^{-4}$&\scriptsize (2.9$\pm$0.1$\pm$2.7)$\times 10^{-4}$  \\ \hline
$\gamma_{n 4}$  &\scriptsize 8$\pm$3$\pm$10                       &\scriptsize  4.95$\pm$0.01$\pm$1.65                    &\scriptsize 5.21$\pm$0.02$\pm$4.75                  &\scriptsize 3.7$\pm$0.1$\pm$2.1                    \\ \hline
$A_{n 6}$       &\scriptsize -                                    &\scriptsize(-4.65$\pm$0.04$\pm$5.07)$\times 10^{-5}$   &\scriptsize(-6.2$\pm$0.1$\pm$5.5)$\times 10^{-5}$   &\scriptsize (-7.5$\pm$0.4$\pm$4.1)$\times 10^{-5}$  \\ \hline
$\gamma_{n 6}$  &\scriptsize -                                    &\scriptsize  3.83$\pm$0.01$\pm$1.58                    &\scriptsize 3.84$\pm$0.02$\pm$2.21                  &\scriptsize 2.5$\pm$0.1$\pm$1.9                    \\ \hline
\end{tabular}
\end{table*}

\begin{table*}
\caption{\label{table:ltw} The extracted leading-twist contribution $\eta_n(Q^2)$
shown in Fig.~\ref{fig:twists}, reported with systematic uncertainties.}
\begin{tabular}{|c|c|c|c|c|} \cline{1-5}
$Q^2~[$(GeV/c)$^2]$ & $\eta_2(Q^2)$x$10^{-1}$ & $\eta_4(Q^2)$x$10^{-2}$ & $\eta_6(Q^2)$x$10^{-3}$ & $\eta_8(Q^2)$x$10^{-3}$ \\ \cline{1-5}
  1.025 & 1.694 $\pm$ 0.004 & 2.54 $\pm$ 0.02 &10.88 $\pm$ 0.20 & 7.97 $\pm$ 0.39 \\ \cline{1-5}
  1.075 & 1.685 $\pm$ 0.004 & 2.47 $\pm$ 0.02 &10.03 $\pm$ 0.18 & 6.72 $\pm$ 0.33 \\ \cline{1-5}
  1.125 & 1.675 $\pm$ 0.003 & 2.40 $\pm$ 0.02 & 9.35 $\pm$ 0.17 & 5.85 $\pm$ 0.28 \\ \cline{1-5}
  1.175 & 1.666 $\pm$ 0.003 & 2.34 $\pm$ 0.02 & 8.80 $\pm$ 0.16 & 5.21 $\pm$ 0.25 \\ \cline{1-5}
  1.225 & 1.657 $\pm$ 0.003 & 2.29 $\pm$ 0.02 & 8.33 $\pm$ 0.15 & 4.71 $\pm$ 0.23 \\ \cline{1-5}
  1.275 & 1.648 $\pm$ 0.003 & 2.24 $\pm$ 0.02 & 7.93 $\pm$ 0.14 & 4.32 $\pm$ 0.21 \\ \cline{1-5}
  1.325 & 1.640 $\pm$ 0.003 & 2.19 $\pm$ 0.02 & 7.59 $\pm$ 0.14 & 4.00 $\pm$ 0.19 \\ \cline{1-5}
  1.375 & 1.632 $\pm$ 0.003 & 2.15 $\pm$ 0.01 & 7.29 $\pm$ 0.13 & 3.74 $\pm$ 0.18 \\ \cline{1-5}
  1.425 & 1.624 $\pm$ 0.002 & 2.11 $\pm$ 0.01 & 7.03 $\pm$ 0.13 & 3.52 $\pm$ 0.17 \\ \cline{1-5}
  1.475 & 1.616 $\pm$ 0.002 & 2.08 $\pm$ 0.01 & 6.80 $\pm$ 0.12 & 3.33 $\pm$ 0.16 \\ \cline{1-5}
  1.525 & 1.609 $\pm$ 0.002 & 2.05 $\pm$ 0.01 & 6.59 $\pm$ 0.12 & 3.17 $\pm$ 0.15 \\ \cline{1-5}
  1.575 & 1.602 $\pm$ 0.002 & 2.02 $\pm$ 0.01 & 6.40 $\pm$ 0.11 & 3.02 $\pm$ 0.14 \\ \cline{1-5}
  1.625 & 1.596 $\pm$ 0.002 & 1.99 $\pm$ 0.01 & 6.23 $\pm$ 0.11 & 2.90 $\pm$ 0.14 \\ \cline{1-5}
  1.675 & 1.590 $\pm$ 0.002 & 1.96 $\pm$ 0.01 & 6.07 $\pm$ 0.11 & 2.79 $\pm$ 0.13 \\ \cline{1-5}
  1.725 & 1.584 $\pm$ 0.002 & 1.94 $\pm$ 0.01 & 5.93 $\pm$ 0.11 & 2.69 $\pm$ 0.13 \\ \cline{1-5}
  1.775 & 1.578 $\pm$ 0.002 & 1.91 $\pm$ 0.01 & 5.80 $\pm$ 0.10 & 2.60 $\pm$ 0.12 \\ \cline{1-5}
  1.825 & 1.572 $\pm$ 0.002 & 1.89 $\pm$ 0.01 & 5.68 $\pm$ 0.10 & 2.51 $\pm$ 0.12 \\ \cline{1-5}
  1.875 & 1.567 $\pm$ 0.002 & 1.87 $\pm$ 0.01 & 5.56 $\pm$ 0.10 & 2.44 $\pm$ 0.12 \\ \cline{1-5}
  1.925 & 1.562 $\pm$ 0.002 & 1.85 $\pm$ 0.01 & 5.46 $\pm$ 0.10 & 2.37 $\pm$ 0.11 \\ \cline{1-5}
  1.975 & 1.557 $\pm$ 0.002 & 1.83 $\pm$ 0.01 & 5.36 $\pm$ 0.09 & 2.31 $\pm$ 0.11 \\ \cline{1-5}
  2.025 & 1.552 $\pm$ 0.002 & 1.81 $\pm$ 0.01 & 5.27 $\pm$ 0.09 & 2.25 $\pm$ 0.11 \\ \cline{1-5}
  2.075 & 1.548 $\pm$ 0.002 & 1.80 $\pm$ 0.01 & 5.18 $\pm$ 0.09 & 2.20 $\pm$ 0.10 \\ \cline{1-5}
  2.125 & 1.543 $\pm$ 0.002 & 1.78 $\pm$ 0.01 & 5.10 $\pm$ 0.09 & 2.15 $\pm$ 0.10 \\ \cline{1-5}
  2.175 & 1.539 $\pm$ 0.002 & 1.76 $\pm$ 0.01 & 5.02 $\pm$ 0.09 & 2.10 $\pm$ 0.10 \\ \cline{1-5}
  2.225 & 1.535 $\pm$ 0.002 & 1.75 $\pm$ 0.01 & 4.95 $\pm$ 0.09 & 2.06 $\pm$ 0.10 \\ \cline{1-5}
  2.275 & 1.532 $\pm$ 0.002 & 1.74 $\pm$ 0.01 & 4.89 $\pm$ 0.09 & 2.02 $\pm$ 0.10 \\ \cline{1-5}
  2.325 & 1.529 $\pm$ 0.002 & 1.72 $\pm$ 0.01 & 4.83 $\pm$ 0.08 & 1.99 $\pm$ 0.09 \\ \cline{1-5}
  2.375 & 1.527 $\pm$ 0.002 & 1.71 $\pm$ 0.01 & 4.78 $\pm$ 0.08 & 1.96 $\pm$ 0.09 \\ \cline{1-5}
\end{tabular}
\end{table*}
\begin{table*}
\begin{tabular}{|c|c|c|c|c|} \cline{1-5}
$Q^2~[$(GeV/c)$^2]$ & $\eta_2(Q^2)$x$10^{-1}$ & $\eta_4(Q^2)$x$10^{-2}$ & $\eta_6(Q^2)$x$10^{-3}$ & $\eta_8(Q^2)$x$10^{-3}$ \\ \cline{1-5}
  2.425 & 1.524 $\pm$ 0.002 & 1.70 $\pm$ 0.01 & 4.73 $\pm$ 0.08 & 1.93 $\pm$ 0.09 \\ \cline{1-5}
  2.475 & 1.522 $\pm$ 0.002 & 1.69 $\pm$ 0.01 & 4.68 $\pm$ 0.08 & 1.90 $\pm$ 0.09 \\ \cline{1-5}
  2.525 & 1.520 $\pm$ 0.002 & 1.68 $\pm$ 0.01 & 4.63 $\pm$ 0.08 & 1.88 $\pm$ 0.09 \\ \cline{1-5}
  2.575 & 1.518 $\pm$ 0.002 & 1.67 $\pm$ 0.01 & 4.59 $\pm$ 0.08 & 1.85 $\pm$ 0.09 \\ \cline{1-5}
  2.625 & 1.516 $\pm$ 0.002 & 1.66 $\pm$ 0.01 & 4.54 $\pm$ 0.08 & 1.83 $\pm$ 0.09 \\ \cline{1-5}
  2.675 & 1.514 $\pm$ 0.002 & 1.65 $\pm$ 0.01 & 4.50 $\pm$ 0.08 & 1.80 $\pm$ 0.08 \\ \cline{1-5}
  2.725 & 1.512 $\pm$ 0.002 & 1.64 $\pm$ 0.01 & 4.46 $\pm$ 0.08 & 1.78 $\pm$ 0.08 \\ \cline{1-5}
  2.775 & 1.510 $\pm$ 0.002 & 1.63 $\pm$ 0.01 & 4.42 $\pm$ 0.08 & 1.76 $\pm$ 0.08 \\ \cline{1-5}
  2.825 & 1.508 $\pm$ 0.002 & 1.62 $\pm$ 0.01 & 4.39 $\pm$ 0.08 & 1.74 $\pm$ 0.08 \\ \cline{1-5}
  2.875 & 1.506 $\pm$ 0.002 & 1.61 $\pm$ 0.01 & 4.35 $\pm$ 0.08 & 1.72 $\pm$ 0.08 \\ \cline{1-5}
  2.925 & 1.504 $\pm$ 0.003 & 1.61 $\pm$ 0.01 & 4.32 $\pm$ 0.08 & 1.70 $\pm$ 0.08 \\ \cline{1-5}
  2.975 & 1.503 $\pm$ 0.003 & 1.60 $\pm$ 0.01 & 4.28 $\pm$ 0.07 & 1.69 $\pm$ 0.08 \\ \cline{1-5}
  3.025 & 1.501 $\pm$ 0.003 & 1.59 $\pm$ 0.01 & 4.25 $\pm$ 0.07 & 1.67 $\pm$ 0.08 \\ \cline{1-5}
  3.075 & 1.499 $\pm$ 0.003 & 1.58 $\pm$ 0.01 & 4.22 $\pm$ 0.07 & 1.65 $\pm$ 0.08 \\ \cline{1-5}
  3.125 & 1.498 $\pm$ 0.003 & 1.58 $\pm$ 0.01 & 4.19 $\pm$ 0.07 & 1.64 $\pm$ 0.08 \\ \cline{1-5}
  3.175 & 1.496 $\pm$ 0.003 & 1.57 $\pm$ 0.01 & 4.16 $\pm$ 0.07 & 1.62 $\pm$ 0.08 \\ \cline{1-5}
  3.225 & 1.495 $\pm$ 0.003 & 1.56 $\pm$ 0.01 & 4.13 $\pm$ 0.07 & 1.61 $\pm$ 0.07 \\ \cline{1-5}
  3.275 & 1.493 $\pm$ 0.003 & 1.56 $\pm$ 0.01 & 4.10 $\pm$ 0.07 & 1.59 $\pm$ 0.07 \\ \cline{1-5}
  3.325 & 1.492 $\pm$ 0.003 & 1.55 $\pm$ 0.01 & 4.08 $\pm$ 0.07 & 1.58 $\pm$ 0.07 \\ \cline{1-5}
  3.375 & 1.490 $\pm$ 0.003 & 1.54 $\pm$ 0.01 & 4.05 $\pm$ 0.07 & 1.56 $\pm$ 0.07 \\ \cline{1-5}
  3.425 & 1.489 $\pm$ 0.003 & 1.54 $\pm$ 0.01 & 4.02 $\pm$ 0.07 & 1.55 $\pm$ 0.07 \\ \cline{1-5}
  3.475 & 1.488 $\pm$ 0.003 & 1.53 $\pm$ 0.01 & 4.00 $\pm$ 0.07 & 1.54 $\pm$ 0.07 \\ \cline{1-5}
  3.525 & 1.486 $\pm$ 0.003 & 1.52 $\pm$ 0.01 & 3.98 $\pm$ 0.07 & 1.53 $\pm$ 0.07 \\ \cline{1-5}
  3.575 & 1.485 $\pm$ 0.003 & 1.52 $\pm$ 0.01 & 3.95 $\pm$ 0.07 & 1.51 $\pm$ 0.07 \\ \cline{1-5}
  3.625 & 1.484 $\pm$ 0.003 & 1.51 $\pm$ 0.01 & 3.93 $\pm$ 0.07 & 1.50 $\pm$ 0.07 \\ \cline{1-5}
  3.675 & 1.482 $\pm$ 0.003 & 1.51 $\pm$ 0.01 & 3.91 $\pm$ 0.07 & 1.49 $\pm$ 0.07 \\ \cline{1-5}
  3.725 & 1.481 $\pm$ 0.003 & 1.50 $\pm$ 0.01 & 3.89 $\pm$ 0.07 & 1.48 $\pm$ 0.07 \\ \cline{1-5}
  3.775 & 1.480 $\pm$ 0.003 & 1.50 $\pm$ 0.01 & 3.87 $\pm$ 0.07 & 1.47 $\pm$ 0.07 \\ \cline{1-5}
  3.825 & 1.479 $\pm$ 0.003 & 1.49 $\pm$ 0.01 & 3.84 $\pm$ 0.07 & 1.46 $\pm$ 0.07 \\ \cline{1-5}
  3.875 & 1.478 $\pm$ 0.003 & 1.49 $\pm$ 0.01 & 3.82 $\pm$ 0.07 & 1.45 $\pm$ 0.07 \\ \cline{1-5}
\end{tabular}
\end{table*}
\begin{table*}
\begin{tabular}{|c|c|c|c|c|} \cline{1-5}
$Q^2~[$(GeV/c)$^2]$ & $\eta_2(Q^2)$x$10^{-1}$ & $\eta_4(Q^2)$x$10^{-2}$ & $\eta_6(Q^2)$x$10^{-3}$ & $\eta_8(Q^2)$x$10^{-3}$ \\ \cline{1-5}
  3.925 & 1.477 $\pm$ 0.003 & 1.48 $\pm$ 0.01 & 3.81 $\pm$ 0.07 & 1.44 $\pm$ 0.07 \\ \cline{1-5}
  3.975 & 1.476 $\pm$ 0.003 & 1.48 $\pm$ 0.01 & 3.79 $\pm$ 0.07 & 1.43 $\pm$ 0.07 \\ \cline{1-5}
  4.025 & 1.474 $\pm$ 0.003 & 1.47 $\pm$ 0.01 & 3.77 $\pm$ 0.07 & 1.42 $\pm$ 0.07 \\ \cline{1-5}
  4.075 & 1.473 $\pm$ 0.003 & 1.47 $\pm$ 0.01 & 3.75 $\pm$ 0.06 & 1.41 $\pm$ 0.07 \\ \cline{1-5}
  4.125 & 1.472 $\pm$ 0.003 & 1.46 $\pm$ 0.01 & 3.73 $\pm$ 0.06 & 1.40 $\pm$ 0.06 \\ \cline{1-5}
  4.175 & 1.471 $\pm$ 0.003 & 1.46 $\pm$ 0.01 & 3.71 $\pm$ 0.06 & 1.40 $\pm$ 0.06 \\ \cline{1-5}
  4.225 & 1.470 $\pm$ 0.003 & 1.45 $\pm$ 0.01 & 3.70 $\pm$ 0.06 & 1.39 $\pm$ 0.06 \\ \cline{1-5}
  4.275 & 1.469 $\pm$ 0.004 & 1.45 $\pm$ 0.01 & 3.68 $\pm$ 0.06 & 1.38 $\pm$ 0.06 \\ \cline{1-5}
  4.325 & 1.468 $\pm$ 0.004 & 1.45 $\pm$ 0.01 & 3.66 $\pm$ 0.06 & 1.37 $\pm$ 0.06 \\ \cline{1-5}
  4.375 & 1.467 $\pm$ 0.004 & 1.44 $\pm$ 0.01 & 3.65 $\pm$ 0.06 & 1.36 $\pm$ 0.06 \\ \cline{1-5}
  4.425 & 1.467 $\pm$ 0.004 & 1.44 $\pm$ 0.01 & 3.63 $\pm$ 0.06 & 1.35 $\pm$ 0.06 \\ \cline{1-5}
  4.475 & 1.466 $\pm$ 0.004 & 1.43 $\pm$ 0.01 & 3.62 $\pm$ 0.06 & 1.35 $\pm$ 0.06 \\ \cline{1-5}
  4.525 & 1.465 $\pm$ 0.004 & 1.43 $\pm$ 0.01 & 3.60 $\pm$ 0.06 & 1.34 $\pm$ 0.06 \\ \cline{1-5}
  4.575 & 1.464 $\pm$ 0.004 & 1.43 $\pm$ 0.01 & 3.59 $\pm$ 0.06 & 1.33 $\pm$ 0.06 \\ \cline{1-5}
  4.625 & 1.463 $\pm$ 0.004 & 1.42 $\pm$ 0.01 & 3.57 $\pm$ 0.06 & 1.33 $\pm$ 0.06 \\ \cline{1-5}
  4.675 & 1.462 $\pm$ 0.004 & 1.42 $\pm$ 0.01 & 3.56 $\pm$ 0.06 & 1.32 $\pm$ 0.06 \\ \cline{1-5}
  4.725 & 1.461 $\pm$ 0.004 & 1.41 $\pm$ 0.01 & 3.54 $\pm$ 0.06 & 1.31 $\pm$ 0.06 \\ \cline{1-5}
  4.775 & 1.460 $\pm$ 0.004 & 1.41 $\pm$ 0.01 & 3.53 $\pm$ 0.06 & 1.31 $\pm$ 0.06 \\ \cline{1-5}
  4.825 & 1.460 $\pm$ 0.004 & 1.41 $\pm$ 0.01 & 3.52 $\pm$ 0.06 & 1.30 $\pm$ 0.06 \\ \cline{1-5}
  4.875 & 1.459 $\pm$ 0.004 & 1.40 $\pm$ 0.01 & 3.50 $\pm$ 0.06 & 1.29 $\pm$ 0.06 \\ \cline{1-5}
  4.925 & 1.458 $\pm$ 0.004 & 1.40 $\pm$ 0.01 & 3.49 $\pm$ 0.06 & 1.29 $\pm$ 0.06 \\ \cline{1-5}
  4.975 & 1.457 $\pm$ 0.004 & 1.40 $\pm$ 0.01 & 3.48 $\pm$ 0.06 & 1.28 $\pm$ 0.06 \\ \cline{1-5}
  5.025 & 1.456 $\pm$ 0.004 & 1.39 $\pm$ 0.01 & 3.47 $\pm$ 0.06 & 1.27 $\pm$ 0.06 \\ \cline{1-5}
  5.075 & 1.456 $\pm$ 0.004 & 1.39 $\pm$ 0.01 & 3.45 $\pm$ 0.06 & 1.27 $\pm$ 0.06 \\ \cline{1-5}
  5.125 & 1.455 $\pm$ 0.004 & 1.39 $\pm$ 0.01 & 3.44 $\pm$ 0.06 & 1.26 $\pm$ 0.06 \\ \cline{1-5}
  5.275 & 1.453 $\pm$ 0.004 & 1.38 $\pm$ 0.01 & 3.41 $\pm$ 0.06 & 1.25 $\pm$ 0.06 \\ \cline{1-5}
  5.325 & 1.452 $\pm$ 0.004 & 1.37 $\pm$ 0.01 & 3.40 $\pm$ 0.06 & 1.24 $\pm$ 0.06 \\ \cline{1-5}
  5.375 & 1.451 $\pm$ 0.004 & 1.37 $\pm$ 0.01 & 3.38 $\pm$ 0.06 & 1.24 $\pm$ 0.06 \\ \cline{1-5}
  5.475 & 1.450 $\pm$ 0.004 & 1.37 $\pm$ 0.01 & 3.36 $\pm$ 0.06 & 1.23 $\pm$ 0.06 \\ \cline{1-5}
  5.525 & 1.449 $\pm$ 0.004 & 1.36 $\pm$ 0.01 & 3.35 $\pm$ 0.06 & 1.22 $\pm$ 0.06 \\ \cline{1-5}
\end{tabular}
\end{table*}
\begin{table*}
\begin{tabular}{|c|c|c|c|c|} \cline{1-5}
$Q^2~[$(GeV/c)$^2]$ & $\eta_2(Q^2)$x$10^{-1}$ & $\eta_4(Q^2)$x$10^{-2}$ & $\eta_6(Q^2)$x$10^{-3}$ & $\eta_8(Q^2)$x$10^{-3}$ \\ \cline{1-5}
  5.625 & 1.448 $\pm$ 0.004 & 1.36 $\pm$ 0.009 & 3.33 $\pm$ 0.06 & 1.21 $\pm$ 0.06 \\ \cline{1-5}
  5.675 & 1.447 $\pm$ 0.004 & 1.35 $\pm$ 0.009 & 3.32 $\pm$ 0.06 & 1.21 $\pm$ 0.05 \\ \cline{1-5}
  5.725 & 1.447 $\pm$ 0.004 & 1.35 $\pm$ 0.009 & 3.31 $\pm$ 0.06 & 1.20 $\pm$ 0.05 \\ \cline{1-5}
  5.955 & 1.444 $\pm$ 0.004 & 1.34 $\pm$ 0.009 & 3.27 $\pm$ 0.06 & 1.18 $\pm$ 0.05 \\ \cline{1-5}
  6.915 & 1.434 $\pm$ 0.005 & 1.30 $\pm$ 0.009 & 3.11 $\pm$ 0.05 & 1.11 $\pm$ 0.05 \\ \cline{1-5}
  7.267 & 1.431 $\pm$ 0.005 & 1.28 $\pm$ 0.008 & 3.06 $\pm$ 0.05 & 1.09 $\pm$ 0.05 \\ \cline{1-5}
  7.630 & 1.427 $\pm$ 0.005 & 1.27 $\pm$ 0.008 & 3.02 $\pm$ 0.05 & 1.07 $\pm$ 0.05 \\ \cline{1-5}
  8.021 & 1.424 $\pm$ 0.005 & 1.26 $\pm$ 0.008 & 2.97 $\pm$ 0.05 & 1.05 $\pm$ 0.05 \\ \cline{1-5}
  8.847 & 1.418 $\pm$ 0.005 & 1.23 $\pm$ 0.008 & 2.89 $\pm$ 0.05 & 1.01 $\pm$ 0.05 \\ \cline{1-5}
  9.775 & 1.413 $\pm$ 0.005 & 1.21 $\pm$ 0.008 & 2.80 $\pm$ 0.05 & 0.97 $\pm$ 0.04 \\ \cline{1-5}
 10.267 & 1.410 $\pm$ 0.005 & 1.20 $\pm$ 0.008 & 2.77 $\pm$ 0.05 & 0.95 $\pm$ 0.04 \\ \cline{1-5}
 10.762 & 1.407 $\pm$ 0.005 & 1.18 $\pm$ 0.008 & 2.73 $\pm$ 0.05 & 0.94 $\pm$ 0.04 \\ \cline{1-5}
 11.344 & 1.405 $\pm$ 0.005 & 1.17 $\pm$ 0.008 & 2.69 $\pm$ 0.05 & 0.92 $\pm$ 0.04 \\ \cline{1-5}
 12.580 & 1.399 $\pm$ 0.006 & 1.15 $\pm$ 0.008 & 2.62 $\pm$ 0.04 & 0.89 $\pm$ 0.04 \\ \cline{1-5}
 13.238 & 1.397 $\pm$ 0.006 & 1.14 $\pm$ 0.007 & 2.58 $\pm$ 0.04 & 0.88 $\pm$ 0.04 \\ \cline{1-5}
 14.689 & 1.392 $\pm$ 0.006 & 1.12 $\pm$ 0.007 & 2.52 $\pm$ 0.04 & 0.85 $\pm$ 0.04 \\ \cline{1-5}
 17.108 & 1.385 $\pm$ 0.006 & 1.09 $\pm$ 0.007 & 2.42 $\pm$ 0.04 & 0.81 $\pm$ 0.04 \\ \cline{1-5}
 19.072 & 1.380 $\pm$ 0.006 & 1.07 $\pm$ 0.007 & 2.36 $\pm$ 0.04 & 0.78 $\pm$ 0.03 \\ \cline{1-5}
 20.108 & 1.378 $\pm$ 0.006 & 1.06 $\pm$ 0.007 & 2.33 $\pm$ 0.04 & 0.77 $\pm$ 0.03 \\ \cline{1-5}
 21.097 & 1.376 $\pm$ 0.006 & 1.05 $\pm$ 0.007 & 2.31 $\pm$ 0.04 & 0.76 $\pm$ 0.03 \\ \cline{1-5}
 24.259 & 1.370 $\pm$ 0.006 & 1.03 $\pm$ 0.007 & 2.23 $\pm$ 0.04 & 0.73 $\pm$ 0.03 \\ \cline{1-5}
 26.680 & 1.369 $\pm$ 0.007 & 1.01 $\pm$ 0.007 & 2.19 $\pm$ 0.04 & 0.71 $\pm$ 0.03 \\ \cline{1-5}
 32.500 & 1.367 $\pm$ 0.007 & 0.99 $\pm$ 0.006 & 2.10 $\pm$ 0.03 & 0.68 $\pm$ 0.03 \\ \cline{1-5}
 34.932 & 1.367 $\pm$ 0.007 & 0.98 $\pm$ 0.006 & 2.07 $\pm$ 0.03 & 0.67 $\pm$ 0.03 \\ \cline{1-5}
 36.750 & 1.366 $\pm$ 0.007 & 0.97 $\pm$ 0.006 & 2.05 $\pm$ 0.03 & 0.66 $\pm$ 0.03 \\ \cline{1-5}
 43.970 & 1.365 $\pm$ 0.007 & 0.94 $\pm$ 0.006 & 1.99 $\pm$ 0.03 & 0.63 $\pm$ 0.03 \\ \cline{1-5}
 47.440 & 1.365 $\pm$ 0.007 & 0.93 $\pm$ 0.006 & 1.96 $\pm$ 0.03 & 0.62 $\pm$ 0.03 \\ \cline{1-5}
 64.270 & 1.363 $\pm$ 0.007 & 0.90 $\pm$ 0.006 & 1.85 $\pm$ 0.03 & 0.58 $\pm$ 0.02 \\ \cline{1-5}
 75.000 & 1.363 $\pm$ 0.007 & 0.88 $\pm$ 0.006 & 1.80 $\pm$ 0.03 & 0.56 $\pm$ 0.02 \\ \cline{1-5}
 86.000 & 1.362 $\pm$ 0.007 & 0.87 $\pm$ 0.006 & 1.76 $\pm$ 0.03 & 0.55 $\pm$ 0.02 \\ \cline{1-5}
 97.690 & 1.362 $\pm$ 0.007 & 0.85 $\pm$ 0.005 & 1.72 $\pm$ 0.03 & 0.53 $\pm$ 0.02 \\ \cline{1-5}
\end{tabular}
\end{table*}

\section{Comparison to Deuteron}\label{sec:Comparison}
Once the separation between the leading and higher-twist terms
is obtained, we can compare them to the corresponding
terms in the deuteron. The deuteron, being a loosely
bound system, is considered here as representing structure function moments
of an almost free proton and neutron.

\begin{figure}
\begin{center}
\includegraphics[bb=1cm 4cm 20cm 23cm, scale=0.4]{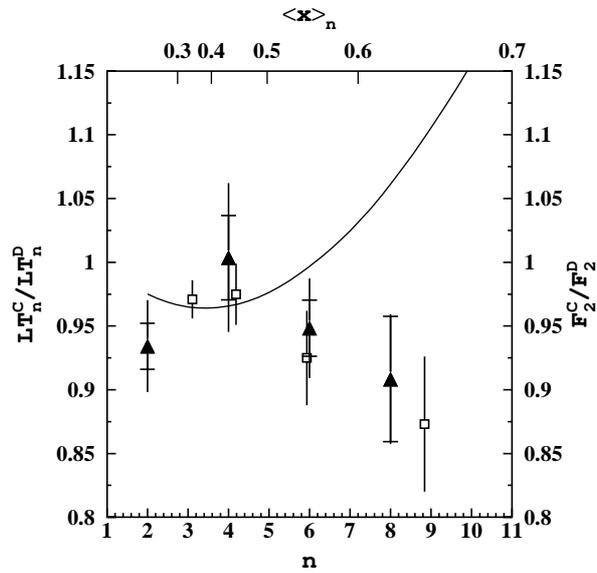}
\caption{\label{fig:emc_lt} The carbon-to-deuteron ratio
of the leading-twist moments (full triangles) compared
with the ratio of corresponding structure functions
obtained in Ref.~\cite{nmc95} (open squares) taken at average $<x>_n$ values
(the cut $Q^2>25$ (GeV/c)$^2$ is applied to avoid higher-twist contamination
in the structure function ratio).
The curve shows the theoretical expectation due to Fermi motion effects.
Internal error bars are statistical only and the total
error bars are the statistical and
systematic uncertainties summed in quadrature.}
\end{center}
\end{figure}

In Fig.~\ref{fig:emc_lt} the ratio of the leading-twist moments
of carbon-to-deuteron is shown with its statistical and systematic
uncertainties. It is compared to the ratio of the corresponding
structure functions obtained in Ref.~\cite{nmc95} and satisfied
the cut $Q^2>25$ (GeV/c)$^2$, ensuring leading-twist dominance.

Considering Fermi motion effects only, and
applying the Impulse Approximation (IA)~\cite{WEST,JAFFE}:
\begin{equation}
F_2^A(x,Q^2)=\int_x^{M_A/M} dz\ f^A(z)\ F_2^N\left(\frac{x}{z},Q^2\right)\ ~,
\label{eq:conv}
\end{equation}
\noindent 
where $f^A(z)$ is the non-relativistic nucleon (light-cone) momentum distribution in the nucleus
and the superscripts $A$ and $N$ are introduced to distinguish the nuclear ($A$) and nucleon ($N$) distributions.
Then, if we define the $n$-th moment of the distribution $f^A(z)$ by:
\begin{equation}
{\cal F}_n^A = \int_0^{M_A/M} dz z^{n-1} f^A(z) ~,
\label{eq:fz_mom}
\end{equation}
\noindent in the moment space, the convolution becomes the product
of the moments~\cite{Buras}:
\begin{equation}
M_n^A(Q^2) = {\cal F}_n^A M_n^N(Q^2)  ~.
\label{eq:MnA}
\end{equation}
\noindent This leads to a simple relation between the carbon and deuteron
moments in the IA:
\begin{equation}
\frac{M_n^C(Q^2)}{M_n^D(Q^2)} = \frac{{\cal F}_n^C}{{\cal F}_n^D}  ~,
\label{eq:RMnA}
\end{equation}
\noindent where the nucleon structure function moments are canceled.
Therefore, in the IA the ratio of nuclear structure function
moments is reduced to the ratio of the nucleon momentum distribution moments.
This ratio, obtained with nuclear wave functions from Refs.~\cite{Simula_nucl,PARIS},
is shown in Fig.~\ref{fig:emc_lt} by the solid line.
The deviation of the data points from the curve at large $n$ is the consequence
of the EMC effect. Indeed, the ratio of the moments is in good agreement with the ratio
of the structure functions taken at $x$ values corresponding
to the average $<x>_n$ for a given $n$ defined as in Ref.~\cite{osipenko_f2n}.
Hence, in the moment space, we confirm the EMC effect discovered previously in $x$-space.

\begin{figure}
\begin{center}
\includegraphics[bb=1cm 4cm 20cm 23cm, scale=0.4]{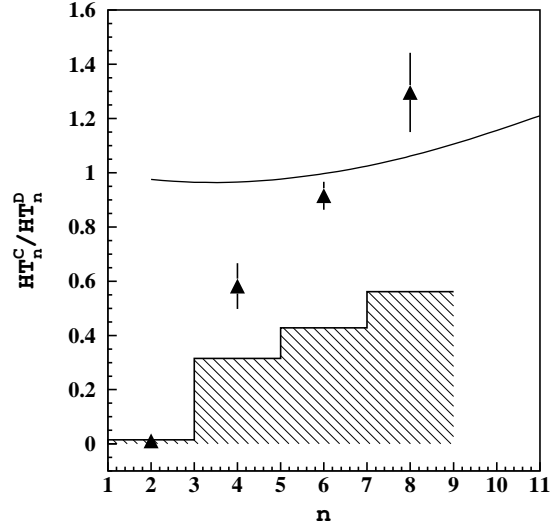}
\caption{\label{fig:emc_ht} The carbon-to-deuteron ratio of the higher-twists moments
at $Q^2=2$ (GeV/c)$^2$.
The curve is the same as in Fig.~\ref{fig:emc_lt}.
The systematic uncertainties are shown by the hatched histogram.}
\end{center}
\end{figure}

\begin{figure}
\begin{center}
\includegraphics[bb=1cm 6cm 20cm 23cm, scale=0.4]{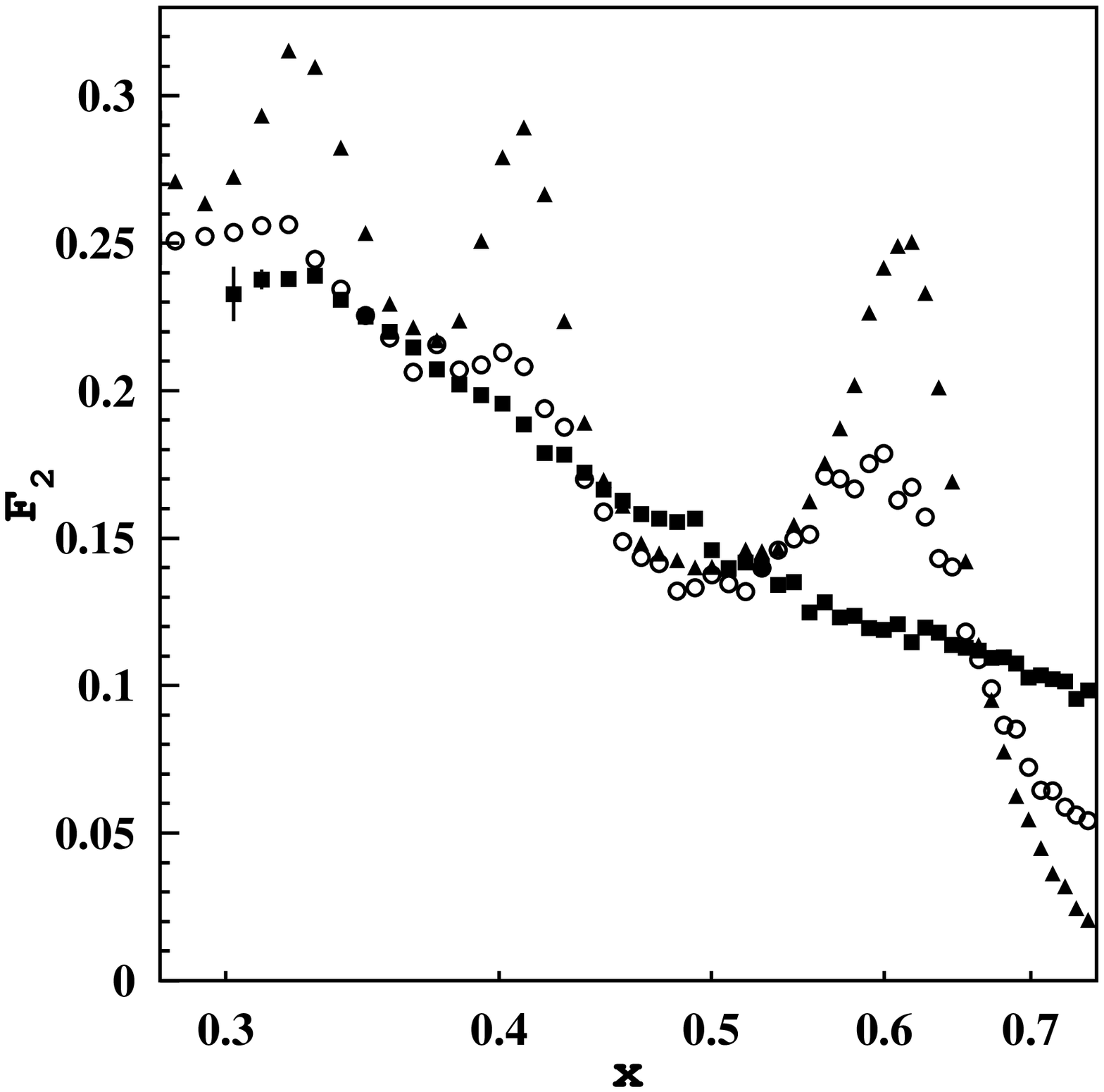}~%
\includegraphics[bb=1cm 6cm 20cm 23cm, scale=0.4]{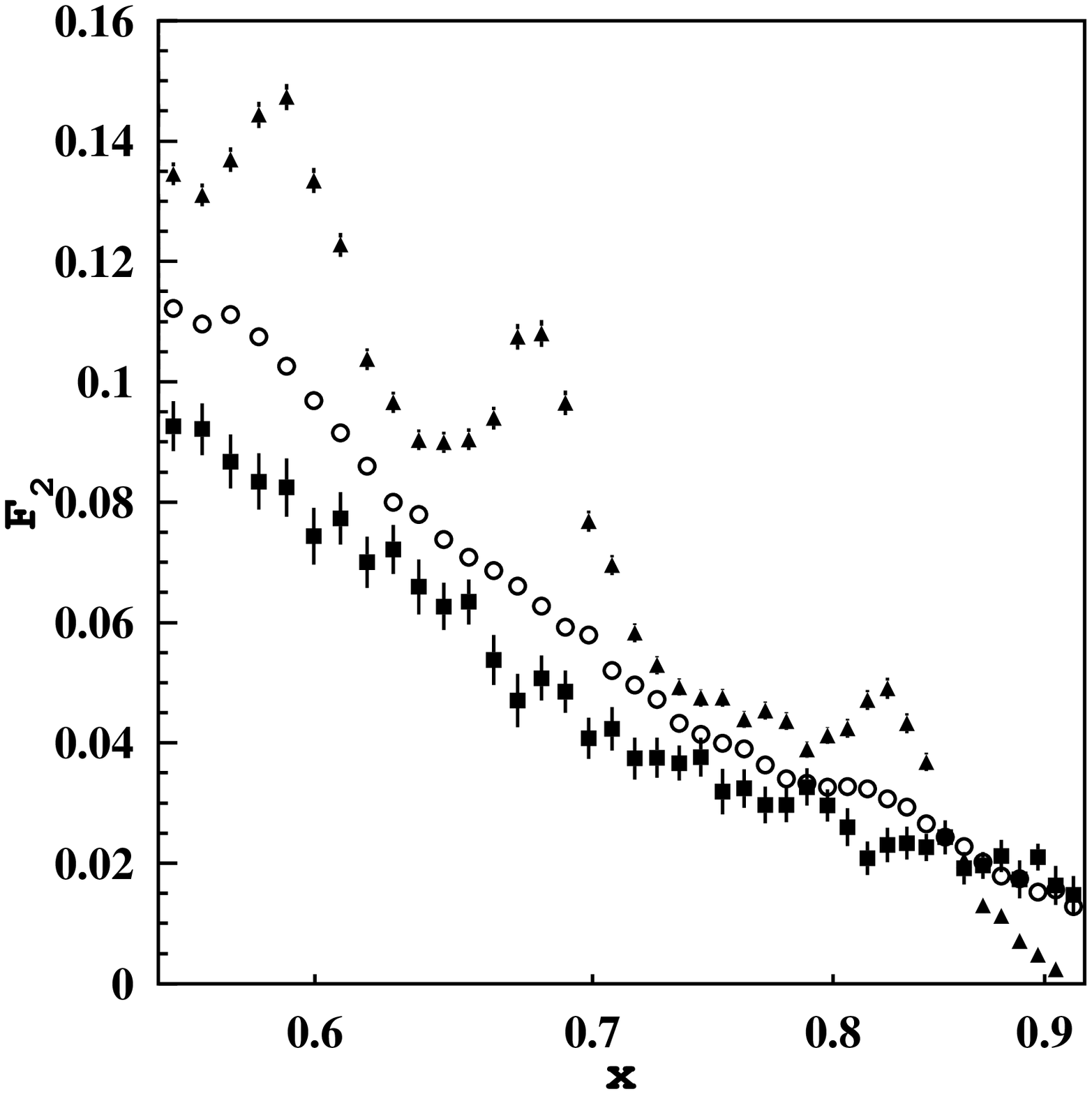}
\caption{\label{fig:F2res}
The nucleon structure function $F_2$ in the resonance region for
proton~\cite{osipenko_f2p} (full triangles), deuteron~\cite{osipenko_f2d} (open circles)
and carbon (full squares) at $Q^2=1$ (GeV/c)$^2$ (left) and $Q^2=3$ (GeV/c)$^2$ (right).
Errors are statistical only.}
\end{center}
\end{figure}

A similar ratio of the carbon to deuteron moments,
but for the total higher-twist contribution
taken at $Q^2=2$ (GeV/c)$^2$, is shown in Fig.~\ref{fig:emc_ht}.
The curve representing the Fermi motion expectation is the same as for the leading-twist case.
The higher-twist ratio is very different from the ratio of leading twists.
Surprisingly, for small $n < 8$, the total higher-twist contribution in complex nuclei
is smaller than in the deuteron. Despite the large systematic uncertainties,
the ratio rises with $n$ almost linearly,
surpassing unity at $n\approx 7$. This behavior is clearly not related
to the Fermi motion of nucleons in the nucleus. Naively one may expect
an overall larger total higher-twist contribution in complex nuclei because
of additional processes, e.g. nuclear FSI and SRC~\cite{Egiyan03,Egiyan06,Science},
which are less important in the deuteron.
A hint of this suppression of the higher twists in nuclei was already observed
in Ref.~\cite{Ricco2}, where the authors tried to connect it to the suppression
of the nucleon excited state form-factors in nuclei~\cite{el_suppress,res_suppress}.
Indeed, looking at
the resonance peaks in the free proton, deuteron and carbon, shown in Fig.~\ref{fig:F2res},
one may conclude that they disappear in nuclei.
However, the suppression factors extracted in Ref.~\cite{res_suppress}
would lead to a much larger difference between the carbon and deuteron
moments. This is because the total resonance contribution
to the structure function moments in the few (GeV/c)$^2$ domain
is very important~\cite{resOPE}. Thus, if resonance form-factors
are really suppressed in nuclei, this has to be compensated
by an almost equivalent increase of the non-resonant contribution to the first four moments.

One explanation of the EMC effect as due to partial quark deconfinement
for a nucleon embedded in nuclear matter was suggested within the rescaling model
proposed in Ref.~\cite{Jaffe_rescaling}.
In Ref.~\cite{Ricco2} it was also connected to the higher-twist suppression in nuclei.
In fact, if one considers the normal QCD potential, partons inside the nucleon
will interact more strongly when approaching the nucleon radius, while partons deep inside the nucleon
will behave as free particles. This intuitive picture is the basis
for the MIT Bag model~\cite{BAGmodel_review}.
The discussion below is a simple-minded speculation suggested by this picture.

\begin{figure}
\begin{center}
\includegraphics[bb=1cm 6cm 20cm 23cm, scale=0.4]{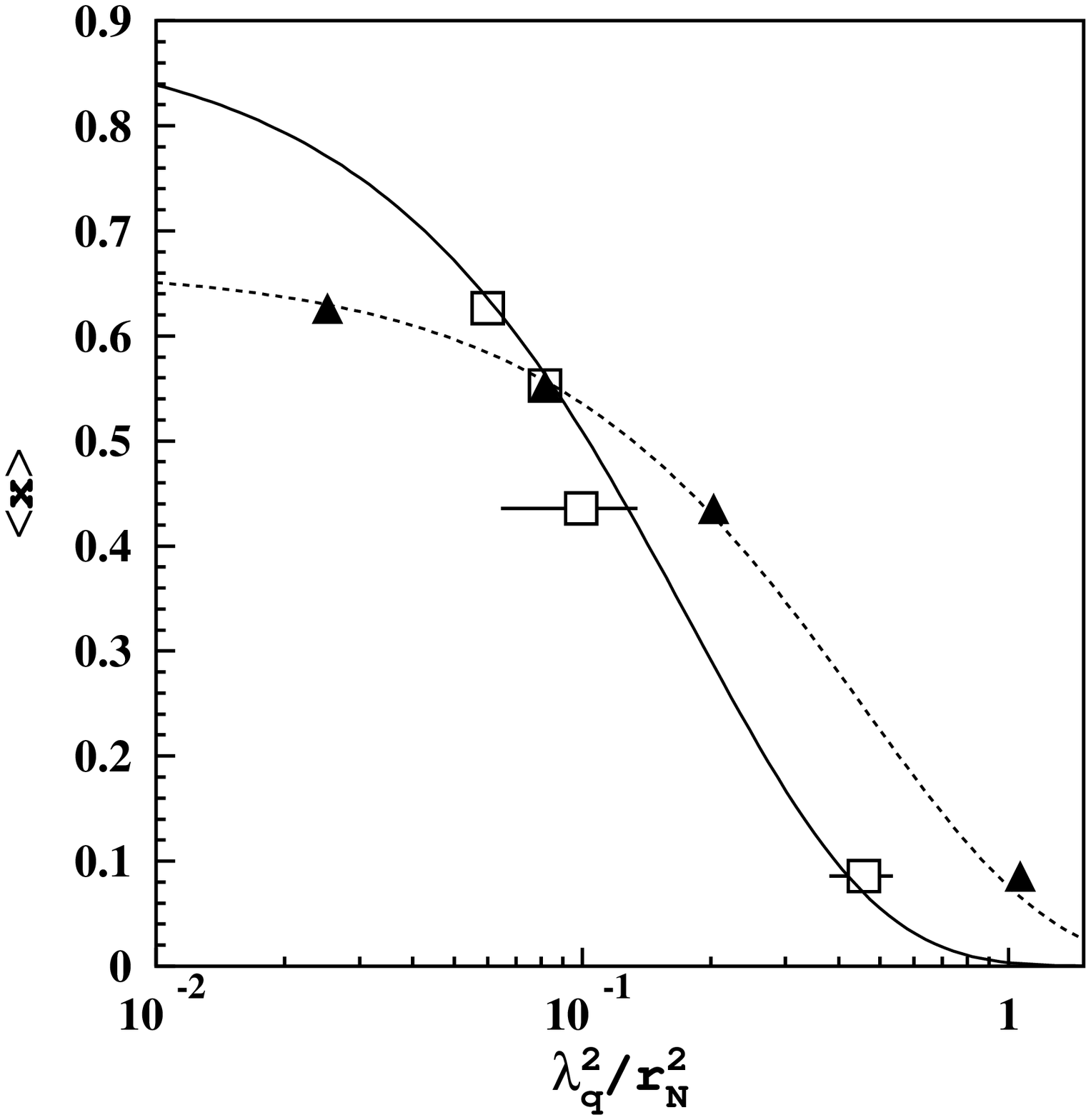}
\caption{\label{fig:MomDistrib} Mean momentum of partons in the nucleon
in the deuteron (open squares) and in carbon (full triangles)
as a function of their mean free path defined in the text (see Eq.~\ref{eq:MFPdef}). The curves show
phenomenological Gaussian parametrizations for the deuteron (solid) and carbon (dashed).
Errors are statistical only.}
\end{center}
\end{figure}

The interaction of the struck parton with the rest of the nucleon will result
in the appearance of higher twists, but as long as the parton is located in the central
region of the nucleon, it will propagate as a free particle, resulting
in the leading-twist dominance.
Therefore, for a given $x$ (or alternatively $n$ in the moment space),
the $Q^2$ value at which the higher-twist contribution becomes significant $Q^2_{HT}$
can be considered as the inverse of the mean free path of the struck parton:
\begin{equation}
\lambda_q \sim r_N \sqrt{\frac{M^2}{Q^2_{HT}}}  ,
\label{eq:MFPdef}
\end{equation}
\noindent where the nucleon charge radius $r_N$ and nucleon mass $M$
define the characteristic scale.
An increase of the nucleon radius
as in the model of Ref.~\cite{Jaffe_rescaling} would lead to a shift of
the higher-twist contribution to lower $Q^2$ for all $n$. Both observed ratios of the leading
and higher twists are compatible with such a naive picture.
One can combine these observations together to visualize
the spatial distribution of the free partons inside the nucleon
as shown in Fig.~\ref{fig:MomDistrib}.
Here we defined $Q^2_{HT}$ used in Eq.~\ref{eq:MFPdef} as the $Q^2$ value at which the total
higher-twist contribution in the given $n$th moment reaches 5\% of the leading twist.
As one can see in the nucleon bound inside the carbon nucleus, partons have
a larger free propagation range,
but the amount of highly energetic free partons is decreased, keeping the integrated
strength almost constant.

\section{Conclusions}\label{sec:Conclusions}
A measurement of inclusive electron scattering by a carbon nucleus
was performed in a wide two-dimensional range of $x$ and $Q^2$.
This measurement improves the kinematic coverage of previously
available data in the low to medium $Q^2$ domain and when combined 
with the existing world data set
allows a calculation of the Nachtmann moments for $n=2,4,6$ and $8$ over
the $Q^2$-range $0.2 - 150$~(GeV/c)$^2$.
The carbon-to-deuteron ratio of the leading-twist $F_2$ moments
exhibits the well known EMC effect by deviating from the Fermi motion
expectation. The deviation is compatible with that observed previously
in $x$ space in DIS.
For the first time we obtained the carbon-to-deuteron ratio of the total higher-twist
contributions to the $F_2$ moments of the nucleon.
Despite large systematic uncertainties, this ratio has a surprising behavior,
increasing almost linearly with the moment order $n$,
and lies well below unity for $n < 7$. This suppression of the higher twists
in the nucleon bound in the nuclear matter cannot be described by
the strong damping of the nucleon excitations
in nuclei suggested in Ref.~\cite{res_suppress}.
The comparison between the relative contributions of higher twists
in carbon and deuteron suggests a wider distribution of free partons in
the bound nucleon. We speculate therefore that the spatial shape of
the nucleon, thought as a bag of free partons, is enhanced when it
is immersed in the nuclear matter, but in contrast to the rescaling model,
the overall probability to find a parton is conserved.

Our analysis indicates a need for further measurements encompassing the kinematic region of 
$Q^2$ from 5 to 40 (GeV/c)$^2$ and large $x$,
and $Q^2>10$ (GeV/c)$^2$ and low $x$.  
The first region can be
explored with 12-GeV Jefferson Lab upgrade~\cite{JLab12GeV}, while 
the second domain would be accessible with the construction of an 
electron-ion collider~\cite{EIC}.

\section{Acknowledgments}
This work was supported by the Istituto Nazionale di Fisica Nucleare,
the French Commissariat \`a l'Energie Atomique, 
the French Centre National de la Recherche Scientifique,
the U.S. Department of Energy, the National Science Foundation and
the National Research Foundation of Korea.
The Southeastern Universities Research
Association (SURA) operated the Thomas Jefferson National Accelerator
Facility for the United States
Department of Energy under contract DE-AC05-84ER40150.

\end{document}